\documentclass[conference]{IEEEtran}

\makeatletter
\def\endthebibliography{%
  \def\@noitemerr{\@latex@warning{Empty `thebibliography' environment}}%
  \endlist
}
\makeatother

\usepackage[cmex10]{amsmath}
\DeclareMathOperator*{\argmax}{arg\,max}
\DeclareMathOperator*{\argmin}{arg\,min}
\usepackage{bm}
\usepackage{amsfonts}
\usepackage{relsize}
\usepackage{cleveref}
\usepackage{cuted}
\usepackage{subfig}
\usepackage{graphicx}
\usepackage{epstopdf}
\usepackage{algorithm}
\usepackage{bbm}
\usepackage{comment}
\usepackage{multirow}
\usepackage{flushend}
\usepackage{booktabs}
\usepackage{textcomp}
\usepackage{cite}
\usepackage{subfig}
\usepackage{xcolor}
\usepackage{enumitem} 

\title{Machine Learning Tips and Tricks for Power Line Communications}

\author{\IEEEauthorblockN{Andrea M. Tonello\thanks{The authors are with the University of Klagenfurt - Chair of Embedded Communication Systems, 9020 Klagenfurt, Austria. (e-mail: \{andrea.tonello, nunzio.letizia, davide.righini, francesco.marcuzzi\}@aau.at)}, \textit{Senior Member, IEEE}, Nunzio A. Letizia, Davide Righini and Francesco Marcuzzi}}

\IEEEoverridecommandlockouts

\begin{document}
\maketitle
\thispagestyle{plain}
\pagestyle{plain}

\begin{abstract}
A great deal of attention has been recently given to Machine Learning (ML) techniques in many different application fields. This paper provides a vision of what ML can do in Power Line Communications (PLC).
We firstly and briefly describe classical formulations of ML, and distinguish deterministic from statistical learning models with relevance to communications.
We then discuss ML applications in PLC for each layer, namely, for characterization and modeling, for the development of physical layer algorithms, for media access control and networking. Finally, other applications of PLC that can benefit from the usage of ML, as grid diagnostics, are analyzed.
Illustrative numerical examples are reported to serve the purpose of validating the ideas and motivate future research endeavors in this stimulating signal/data processing field.
\end{abstract}


\maketitle
\section{Introduction}
\label{sec:introduction}
Modern communication systems have reached a high degree of performance, meeting demanding requirements in numerous application fields. Significant progress in the analysis and design of communication systems has been rendered possible by the milestone work of C. Shannon \cite{Shannon1948} that provided a methodological approach to attack the challenge of reliably transmitting information through a given communication mean. Shannon's mathematical approach suggests to represent the system as a chain of blocks mathematically modeled, namely, the transmitter, the channel, and the receiver. The transmitter can be further divided into a source coder, a channel coder and a signal modulator. The channel is represented by a transfer function (in most cases considered linear time invariant, or time variant) and an additive noise term. Three generations of scientists and engineers grew up with this mathematical mindset which provided tools to acquire domain knowledge and use it to build a model for each block, so that the overall behavior becomes known. Such a framework intrinsically has the advantage that each block can be individually studied and optimized. We would refer to this approach as \textit{physical} and \textit{bottom-up}.

From an epistemological point of view, the \textit{mathematical theory of communications} is based on knowledge coming from \textit{a priori} justifications and relying on intuitions and the nature of these intuitions, which is intrinsically what mathematics does. On the contrary, \textit{a posteriori} knowledge is created by what is known from experience, therefore generated afterwards with an \textit{empirical} and \textit{top-down} approach. The immense contributions of I. Kant with his philosophy of transcendental aesthetics and logic \cite{Kant}, made a step further into the understanding and the definition of knowledge: knowledge of the structure of time and space and their relationships, is \textit{a priori} knowledge; knowledge acquired from observations is \textit{a posteriori} knowledge; most of our knowledge comes from the process of learning and observing phenomena, and without a priori knowledge it is impossible to reach the true knowledge. In this respect, Machine Learning (ML) \cite{Bishop2006},\cite{DLBook}, which is the topic of this paper, can be considered an implementation by humans of techniques in machines to acquire knowledge from \textit{a posteriori} observations of natural phenomena. The origin of success of ML relies on its ability to derive relations among phenomena and potentially discover the hidden (latent) state of a system, i.e, potentially provide an intrinsic \textit{true} knowledge of the system. System identification and model based design through the aid of ML \cite{Murphy2012} constitute a first step to find undiscovered system properties via a mixed \textit{a priori - a posteriori} learning approach, which, retrospectively, follows Kant's philosophical structure.

ML is indeed bringing new lymph in the domain of communication systems modeling, design, optimization, and management. It provides a paradigm shift: rather than concentrating on a physical bottom-up description of the communication scheme, ML aims to learn and capture information from a collection of data, to derive the input-output relations of the system, or of a given task in the system. We would argue that a miraculous solution of communications challenges with ML does not yet exist. In addition, what is learned via ML tools is not necessarily representative of the physical reality, i.e., wrong believes about relations and dependencies among data may be generated. Consequently, the results have to be validated through the support of a probabilistic approach and an understanding of the system physics. But the path has been mapped out: ML offers a great deal of opportunities to research and design communication systems.

\subsection{But what are the domains of application of ML in communications?}
The applications of ML in communications are a multitude and cover all three fundamental protocol stacks: the physical layer, the MAC layer and the network layer. More specifically, the same applies to Power Line Communications (PLC) which is the technology that exploits the existing power deliver infrastructure to convey information signals \cite{lampe2016power}. When things become complex and a bottom-up model is difficult to derive or has too many uncertainties, ML can help, no doubt. This is particularly true in PLC since, despite the advances in channel and noise modeling \cite{PLCbookch2}, the communication media is still not fully understood and modeled especially when it comes to noise and interference. Consequently, the transceiver techniques designed so far might not be optimal \cite{PLCbookch5}. Media access control and resource allocation in massive PLC networks (as smart metering ones) is an extremely complex task that can benefit from ML approaches \cite{PLCbookch6}. The analysis of PLC signals exchanged among nodes can reveal properties of the grid status and detect anomalies in the cables, loads and generators, which is relevant for grid predictive maintenance \cite{6338332,AfricaTonello}.

\subsection{Paper Contribution}
In this paper, we will discuss in detail the application of ML in PLC providing concrete insights (tips) of what can be done and with what ML tools (tricks). Several numerical examples are reported to validate the ideas and to stimulate further work in this research domain. To better understand ML, we start our short journey into ML for PLC by proving a compact introduction to ML with focus on applications in communications. This serves also the purpose of surveying, at the best of the authors knowledge, the existing literature on the topic and the initial studies conducted.

In detail, the paper is organized as follows. In Sec. II, ML fundamentals for both supervised and unsupervised learning are reported. Specific tools are described. They include artificial neural networks, and support vector machines (for supervised ML), and clustering, autoencoders, and generative networks (for unsupervised ML). Convolutional and Recurrent Neural Networks as well as Reinforcement Learning are also briefly discussed. In Sec. III, we focus on PLC and ML for the characterization of the medium and its modeling through the use of a data driven, synthetic approach. In Sec. IV, ML for physical layer PLC is discussed while the MAC and network layers are considered in Sec. V. Other applications of ML for PLC are the topic of Sec. VI. The conclusions then follow.

\section{Machine Learning Basics for Communications}
Following the definition provided by Mitchell \cite{Mitchell1997}, ML algorithms can be categorized according to the learning process (the kind of experience $E$ the machine has), the specific task $T$ and the performance measure $P$. If we focus on the experience, learning is divided into \textit{supervised, unsupervised} and \textit{reinforcement} learning. According to different interpretations of the experience, one could distinguish between a \textit{deterministic} or \textit{probabilistic} interpretation/description of the system. Since in communication theory, signals are always studied as stochastic processes, we prefer to encapsulate ML tools into \textit{statistical learning} tools, thus, under a stochastic/probabilistic framework.

Now, as discussed in the introduction, the paradigm shift in ML for communications is to frame the problem into an initial black-box description of the system and then learn and capture information from a collection of data so that a target task can be realized \cite{Simeone2018}. Examples of tasks, are system modeling, or more in general system identification, data detection, resource allocation, network management etc.

In the following, we summarize learning methods and tools and we report specific examples of ML based solutions already proposed in the literature using however a unified description approach.

\begin{table*}
	\scriptsize 
	\centering
	\begin{minipage}{\textwidth}\centering
		\centering
		\includegraphics[scale = 0.6]{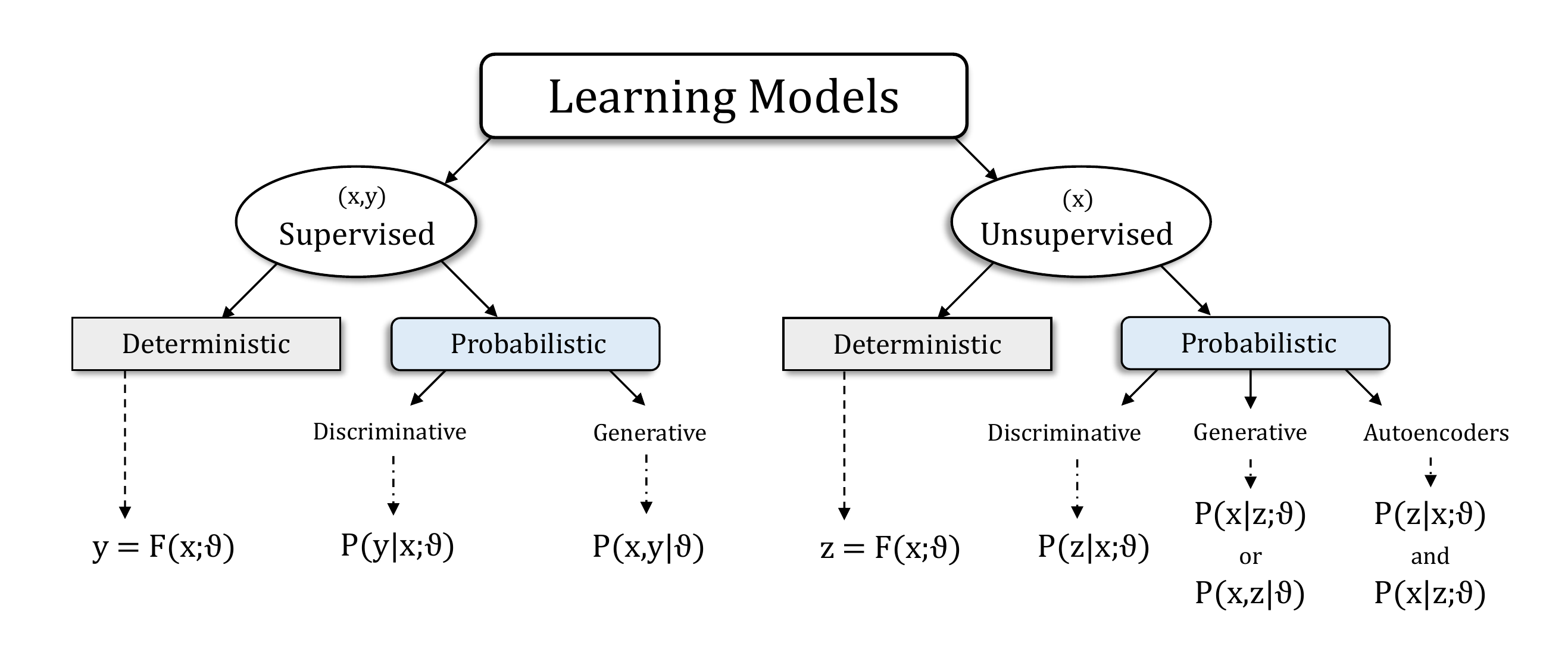}
		\captionof{figure}{Taxonomy of learning models.}
		\label{LearningModels}
		\noindent\makebox[\linewidth]{\rule{0.85\paperwidth}{0.4pt}} 
	\end{minipage}
\end{table*}
\subsection{Supervised Learning}
\label{sec:SL}
\subsubsection{Preliminaries and Definitions}

Let $(\mathbf{x}_i,\mathbf{y}_i)\sim p(\mathbf{x},\mathbf{y})$, $i=1,\dots, N$, be samples collected into a training set $\mathcal{D}$ belonging to the joint distribution (pdf) $p(\mathbf{x},\mathbf{y})$. Supervised learning, under a deterministic model, aims to find a mapping between all pairs of input-output vectors $(\mathbf{x}_i,\mathbf{y}_i)$, thus, an inferred function $F$ that element-wise satisfies $\mathbf{y}=F(\mathbf{x})$. The ideal scenario would map unseen samples $\mathbf{\tilde{x}}$ into the right, initially unknown, label/target $\mathbf{\tilde{y}}$. As we want to address the problem under a stochastic/probabilistic approach, we state that probabilistic supervised learning tries to predict $\mathbf{y}$ from $\mathbf{x}$ by estimating $p(\mathbf{y}|\mathbf{x})$ under a \textit{discriminative model} or by estimating the joint distribution $p(\mathbf{x},\mathbf{y})$ under a \textit{generative model}. Fig. \ref{LearningModels} schematically distinguishes between deterministic and probabilistic approaches in ML.

If the outputs are continuous variables, we consider it as a \textit{regression} problem, while if the targets are discrete, then we have a \textit{classification} problem.
A standard way to proceed during the learning process is to define a cost function $C$, namely a performance measure that evaluates the quality of our prediction $\mathbf{\hat{y}}$. In most applications, we can rely only on the observed dataset $\mathcal{D}$ and derive an empirical sample distribution since we do not have knowledge of the true joint distribution $p(\mathbf{x},\mathbf{y})$. In particular, the goal of the training process is to minimize
\begin{equation}
\label{GeneralCost}
C(\mathbf{\hat{y}}) = \mathbb{E}_{(\mathbf{x},\mathbf{y}) \sim \mathcal{D}}[\delta(\mathbf{y},\mathbf{\hat{y}})]
\end{equation}
where $\delta$ is a measure of distance between the wanted target $\mathbf{y}$ and the prediction $\mathbf{\hat{y}}$, and $\mathbb{E}$ denotes expectation.

\subsubsection{Tools: Neural Networks}
\label{subsec:NN}
Neural Networks (NNs) are among the most popular tools in this field since they are known being universal function approximators \cite{Hornik1989}, they can be implemented in parallel on concurrent architectures and most importantly, they can be trained by backpropagation \cite{Rumelhart1985}.

\begin{figure}
\centering
\includegraphics[scale = 0.33]{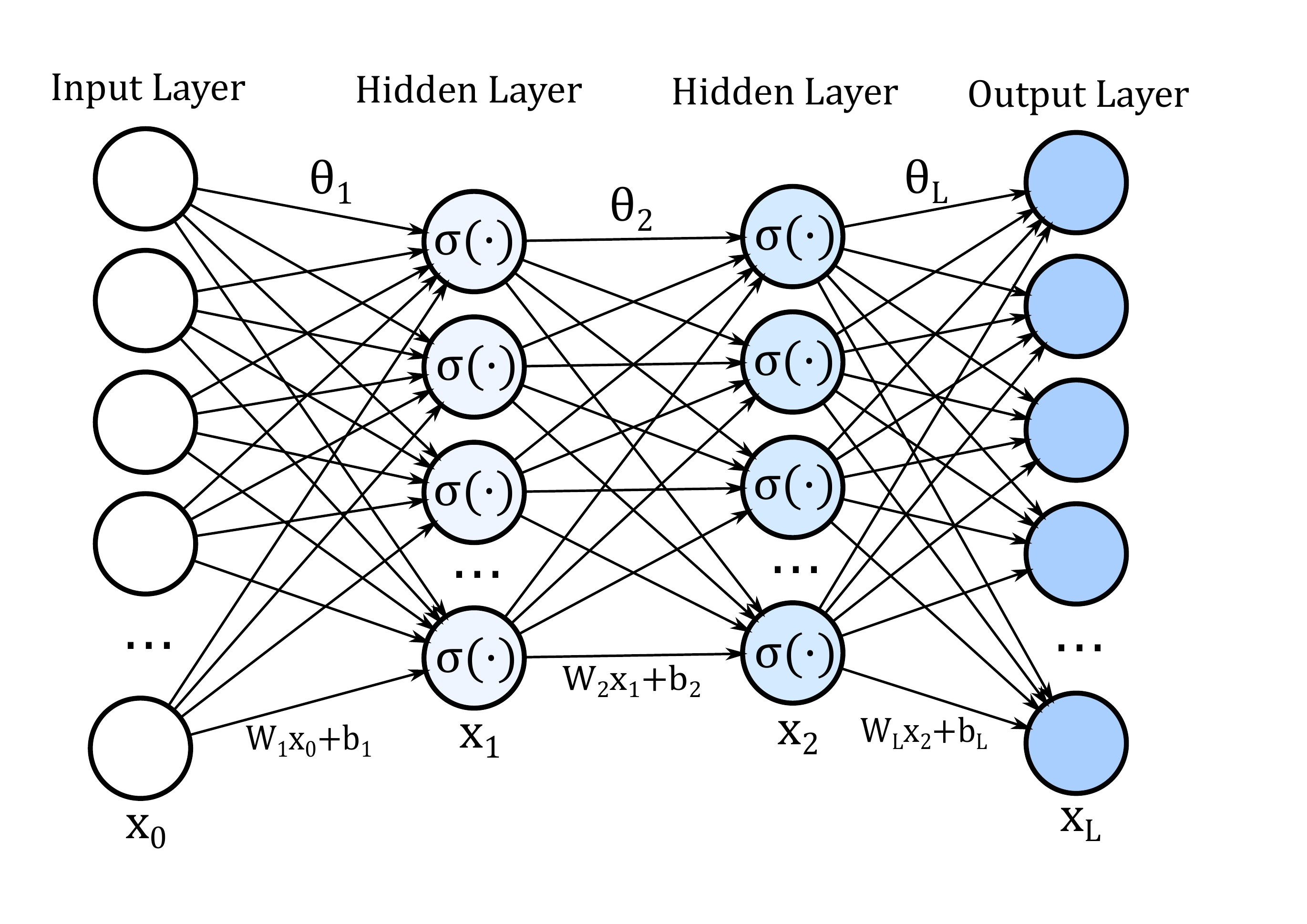}
\captionof{figure}{Structure of a fully connected neural network with $2$ hidden layers.}
\label{NeuralNetwork}
\end{figure}

A feedforward neural network with $L$ layers maps a given input $\mathbf{x}_0 \in \mathbb{R}^{D_0}$ to an output $\mathbf{x}_L \in \mathbb{R}^{D_L}$ by implementing a function $F(\mathbf{x}_0;\mathbf{\theta})$ where $\mathbf{\theta}$ represents the parameters of the NN. To do so, the input is processed through $L$ iterative steps
\begin{equation}
\label{Layer}
\mathbf{x}_l = f_l(\mathbf{x}_{l-1};\theta_l), \; l=1,\dots, L
\end{equation}
where $f_l(\mathbf{x}_{l-1};\theta_l)$ maps the input of the $l$-th layer to its output. The most used layer is the fully connected one, whose mapping is expressed as
\begin{equation}
\label{Layer}
f_l(\mathbf{x}_{l-1};\theta_l) = \sigma(\mathbf{W}_l\cdot \mathbf{x}_{l-1}+\mathbf{b}_l)
\end{equation}
where $\sigma(\cdot)$ is the activation function while $\mathbf{W}_l$ and $\mathbf{b}_l$ are the parameters, weights and the biases, respectively.
According to the specific application, several different types of layers and activation functions can be defined. Fig. \ref{NeuralNetwork} shows a general fully-connected architecture.

Defined a metric $\delta$ and a cost function $C$, the easiest and most classical algorithm to find the feasible set of parameters $\mathbf{\theta}$ is the gradient descent which iteratively updates $\mathbf{\theta}$ as $\mathbf{\theta}_t = \mathbf{\theta}_{t-1}-\eta\nabla C(\mathbf{\theta}_{t-1})$ where $\eta$ is the learning rate. Its popular variants are Stochastic Gradient Descent (SGD) and adaptive learning rates (Adam) \cite{AdamKingma}.
Common choices for the cost function, in a deterministic model, are the mean squared error and categorical cross entropy, for which $\delta$ in \eqref{GeneralCost} takes the form $\delta(\mathbf{y},\mathbf{\hat{y}}) = ||\mathbf{y}-\mathbf{\hat{y}}||^2$ and $\delta(\mathbf{y},\mathbf{\hat{y}}) = -\mathbf{y}\cdot \log{\mathbf{\hat{y}}}$, respectively.

In the context of probabilistic discriminative models, the fundamental learning criterion is the \textit{maximum likelihood} estimator which finds a value of $\theta$ in a parameterized family of models $p(\mathbf{y}|\mathbf{x};\theta)$ that is the most likely to have generated the observed data $\mathcal{D}$, formally
\begin{equation}
\mathbf{\theta} = \argmax_{\mathbf{\theta}}p(\mathbf{y}|\mathbf{x};\theta).
\label{ML}
\end{equation}
From the dataset samples and using the log-likelihood function, \eqref{ML} can be empirically estimated as
\begin{equation}
\mathbf{\theta} = \argmax_{\mathbf{\theta}} \sum_{i=1}^{N}{\log p(\mathbf{y}_i|\mathbf{x}_i;\theta)}.
\label{logML}
\end{equation}
Note that the deterministic and probabilistic approaches collapse together when we consider Gaussian or categorical models since we will end up solving a least squares or a cross-entropy minimization problem, respectively.
We leave the description of generative models for Sec. \ref{sec:UL} since they are typically and formally introduced under unsupervised learning, and then they have been extended to supervised versions. This is the reason why they can be cast in hybrid models.

NNs are only one of the several different tools introduced so far by the ML's community. Since they are excellent in finding patterns, in the context of PLC, as we will discuss, they can be used to find deterministic and probabilistic relationships between physical measured quantities such as the line impedence, the channel transfer function, the radiated power. Channel emulation can be realized with neural networks, an early example of which is \cite{Ma2008}. NNs can also be used for prediction and network management purposes.

\subsubsection{Tools: Support Vector Machines}
\label{subsec:SVM}
\begin{figure}
\centering
\includegraphics[scale = 0.25]{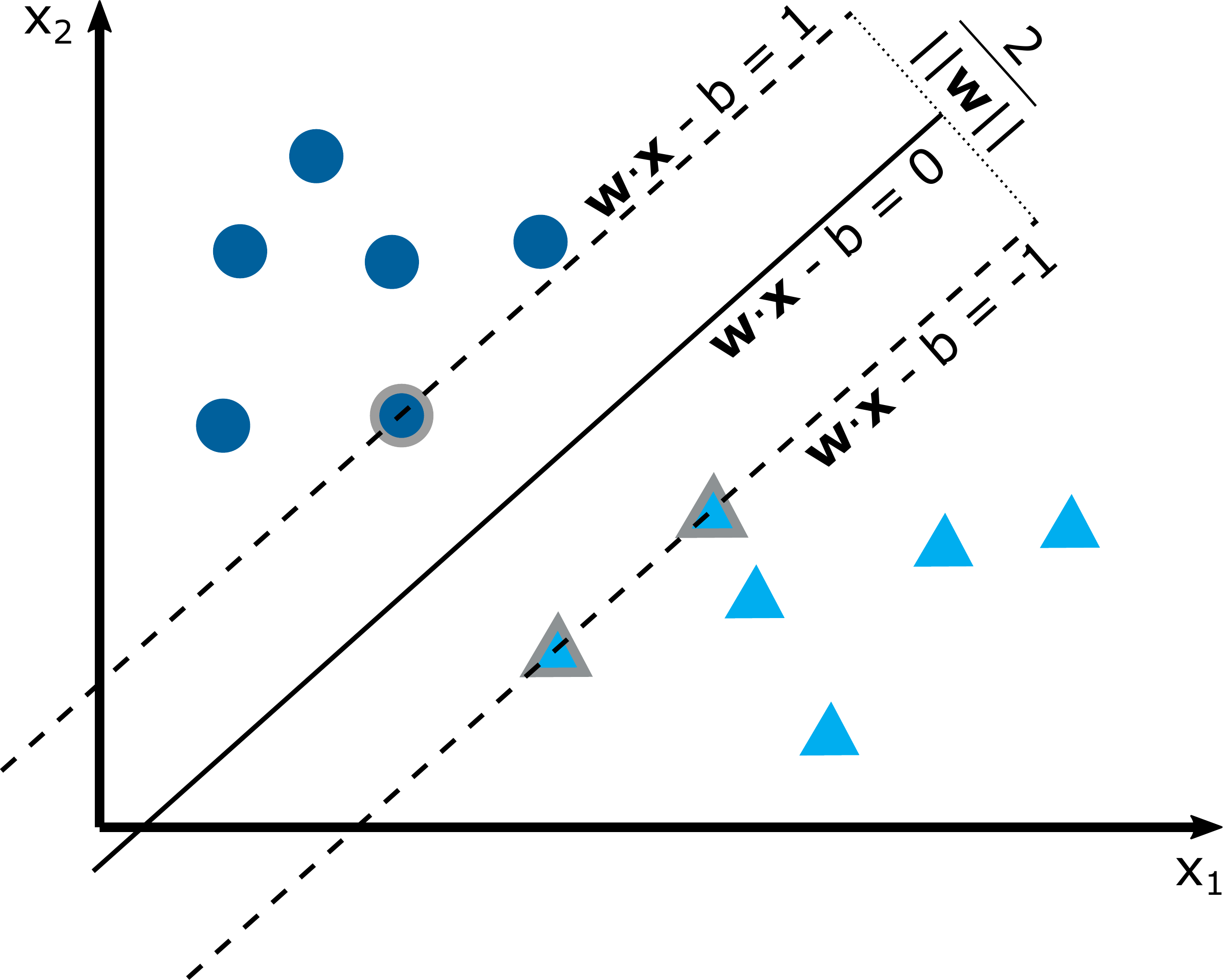}
\captionof{figure}{Decision boundaries and maximum separation for SVM framework.}
\label{SVM}
\end{figure}

Another relevant approach to supervised learning is the usage of a Support Vector Machine (SVM) \cite{Boser1992},\cite{Cortes1995}. We briefly describe the SVM classifier since regression follows from the same idea with minor modifications. The SVM classifier predicts that a certain sample $\mathbf{x}_i$ belongs to a class $y_i$ if $\mathbf{w}^T \mathbf{x}_i-b$ is positive, where $\mathbf{w}^T \mathbf{x}_i-b=0$ is the equation for the decision boundary. The aim is to maximize the distance $d=2/||\mathbf{w}||$ between the supporting rescaled hyperplanes $\mathbf{w}^T \mathbf{x}_i-b=1$ and $\mathbf{w}^T \mathbf{x}_i-b=-1$ (see Fig \ref{SVM}). This leads to the following dual Lagrangian formulation for the linear SVM
\begin{equation}
\mathcal{L}_d = \sum_{i=1}^{N}{\alpha_i}-\frac{1}{2}\sum_{i=1}^{N}{\sum_{j=1}^{N}{\alpha_i \alpha_j y_i y_j \mathbf{x}_i \mathbf{x}_j}}
\end{equation}
where $\alpha_i$ are the Lagrange multipliers (non-negatives) and $N$ the number of samples. When data are non linearly separable in the domain space, transforming them into an higher dimensional one using the mapping $\mathbf{x} \rightarrow \phi(\mathbf{x})$ increases the complexity of the classifier. A concept called \textit{kernel trick} overcomes this problem; the key idea behind the kernel trick is that the inner product between $\mathbf{x}_i$ and $\mathbf{x}_j$ is similar to the inner product between $\phi(\mathbf{x}_i)$ and $\phi(\mathbf{x}_j)$. Moreover, since we are not interested in knowing $\phi$ but only the scalar product $\mathbf{x}_i \cdot \mathbf{x}_j$, such product is represented by the kernel function $K(\mathbf{x}_i,\mathbf{x}_j)$. Consequently, we can make predictions using the function
\begin{equation}
f(\mathbf{x}) = b + \sum_i \alpha_i K(\mathbf{x},\mathbf{x}_i),
\end{equation}
which is a non-linear function w.r.t. $\mathbf{x}$ but linear w.r.t. $\phi(\mathbf{x})$ and $\alpha$. The most commonly used kernel is the radial basis function (RBF) kernel with expression
\begin{equation}
K(\mathbf{x}_i,\mathbf{x}_j) = \exp\biggl(-\frac{||\mathbf{x}_i-\mathbf{x}_j||^2}{\sigma^2}\biggr).
\end{equation}
As an example, in the context of PLC, SVM was recently explored to monitor and detect cable degradations \cite{Huo2019}.

\subsubsection{Other Tools}
Another very simple non-probabilistic supervised learning algorithm is $k$-nearest neighbors. It is a non-parametric algorithm in the sense that, since there is no learning process, at the test time we predict $y_i$ for a given test input $\mathbf{x}_i$. The main idea is to find the $k$-nearest neighbors (for example using an Euclidean metric) to $\mathbf{x}_i$ in the training set $\mathbf{x}$. The algorithm returns the average, if regression is considered, and the mode, if classification is considered, of the corresponding $y$, also called training class. This algorithm is very easy to implement but it becomes significantly slower as the number of samples increases.

Naive Bayes, decision trees, and random forest \cite{TinKam1995} are other popular examples of supervised learning algorithms.

\subsubsection{Application of Supervised Learning to Communications}
Supervised learning algorithms have been applied in communications scenarios with promising results with focus from physical layer to networking at the edge or at the cloud segment.
\begin{itemize}
\item \textit{Physical Layer}: At the transmitter, the problem of maximizing energy efficiency has been addressed via resource allocation with deep neural networks \cite{Zappone}. At the receiver, several problems have been considered such as data detection and decoding \cite{Gruber2017},  channel estimation and symbol detection in an OFDM system \cite{Ye2018}, linear codes decoding \cite{Nachmani2018}, decoding for non-linear channels like those in optical communications \cite{Argyris2018}, modulation recognition \cite{West2017}. In addition, also radio-localization using discriminant adaptive neural networks has been proposed in \cite{FangLocalization};
\item \textit{Link and Medium Access Control Layer}: feedback on the decodability of the received signal has been proposed using ML techniques in \cite{Strodthoff}; spectrum sensing and allocation in cognitive networks in presence of interfering devices has been done in \cite{Tumuluru}; interference management was considered in \cite{Sun2018} and estimation of the number of active nodes in a wireless network testing various ML techniques was the objective of \cite{DelTesta}.
\item \textit{Network and Application Layers}: At the edge, caching popular contents in echo state networks exploiting a ML framework was investigated in \cite{Chen}; at the cloud, a network controller was designed for routing using a predictive approach \cite{WangRouting}; lastly, a survey of ML techniques for traffic classification was offered in \cite{Nguyen}.
\end{itemize}

\subsection{Unsupervised Learning}
\label{sec:UL}
\subsubsection{Preliminaries and Definitions}
Let $\mathbf{x}_i\sim p(\mathbf{x})$, $i=1,\dots, N$, be samples collected into a training set $\mathcal{D}$ belonging to the probability distribution function (PDF) $p(\mathbf{x})$.
Unsupervised learning aims to find useful properties of the structure of a dataset $\mathcal{D}$, ideally inferring the true unknown distribution $p(\mathbf{x})$.
Several different tasks are solved using unsupervised learning, roles such as: clustering, which divides the data into cluster of similar samples, feature extraction, which transforms data in a different latent space easier to handle and interpret, density estimation and generation/synthesis of new samples, whose objective is to learn, from data in $\mathcal{D}$, the distribution $p(\mathbf{x})$ and to produce new unseen samples from it.
Contrarily to the supervised one, unsupervised learning  does not have an unified accepted formulation. This is even clearer if we think that many unsupervised learning tasks require the introduction or the presence of an hidden variable $\mathbf{z}_i$ for each sample, leading to the selection of different models under a probabilistic approach (as shown in Fig. \ref{LearningModels}):
\begin{itemize}
\item \textit{Discriminative models}, where the latent code $\mathbf{z}_i$ has to be extracted from data $\mathbf{x}_i$ by defining a family $p(\mathbf{z|x};\theta)$ parameterized by a vector $\theta$.
\item \textit{Autoencoders}, where $\mathbf{x}_i$ is encoded into a latent variable $\mathbf{z}_i$ so that recovering $\mathbf{x}_i$ from $\mathbf{z}_i$ is possible through a decoder. This model, as discussed in Sec. \ref{subsec:autoencoders}, is very popular nowadays and can be interpreted also under a supervised learning approach, where $\mathbf{x}_i$ is the input data, and $\mathbf{x}_i$ is itself the corresponding target. In the classical deterministic approach, the problem consists of the parameterization of two functions, namely the encoder $\mathbf{z} = F(\mathbf{x};\theta)$, and the decoder $\mathbf{x} = G(\mathbf{z};\theta)$.
In the probabilistic approach, the encoder has to model $p(\mathbf{z|x};\theta)$, while the decoder models $p(\mathbf{x|z};\theta)$.
\item \textit{Generative models}, where there exists an hidden $\mathbf{z}_i$ that generates the observation $\mathbf{x}_i$. After a specification of a parameterized family $p(\mathbf{z}|\theta)$, the distribution of the observation can be rewritten as $p(\mathbf{x|\theta})=\sum_z p(\mathbf{z|\theta})p(\mathbf{x|z;\theta})$.
\end{itemize}
For a fully and exhaustive description of the different models and the corresponding tools like Expectation Maximization (EM) and Evidence Lower BOund (ELBO), the interested reader is referred to \cite{SimeoneDL}. We will focus only on clustering algorithms for discriminative models and autoencoders since they are now getting a lot of attention by communication researchers.

\subsubsection{Tools: Clustering}
Clustering analysis includes different algorithms whose common task is to discover the hidden unknown structure and relationships between input data. Several algorithms are available in the literature for this purpose.
The most common and reliable ones are K-means, Hierarchical Clustering (HC), clustering using representatives (CURE), and Self Organizing Maps (SOM).

K-means is a clustering algorithm that allows to partition the input data in K sets, based on a certain metric exploiting expectation maximization algorithms.
K-means clustering, considers a set of $N$ data points in a $D$-dimensional space $\mathbb{R}^{D}$ and an integer $K$. The problem is to determine a set of $K$ points in $\mathbb{R}^{D}$, called centroids, so as to minimize the mean squared distance from each data point to its nearest centroid \cite{1017616}.
Usually, this clustering algorithm is solved efficiently with heuristic methods such as Lloyd's.

HC algorithms are mainly classified into agglomerative methods (bottom-up methods) and divisive methods (top-down methods), depending on how the hierarchical dataset is formed \cite{Roux2018}. Multiple implementations and variations of the HC algorithms are described in the literature \cite{4061364}. Essentially, they follow a procedure where a tree of clusters called dendrogram is generated.

The CURE algorithm attempts to uncover the cluster shape using a collection of representative points from the main dataset \cite{leskovec_rajaraman_ullman_2014}. This algorithm is efficient for large databases, and compared to K-means clustering is more robust to outliers and able to identify clusters with complex shapes.

SOM is a grid of map units \cite{846731}. Each unit is represented by a prototype vector that defines its space position. The units are connected to the adjacent ones to form a network. The number of map units corresponds to the final number of clusters, thus, a higher number of units corresponds to a higher accuracy of data separation. A training procedure is used to stretch the network and map the space of input data.

\subsubsection{Tools: Autoencoders}
\label{subsec:autoencoders}
\begin{figure}
\centering
\includegraphics[scale = 0.25]{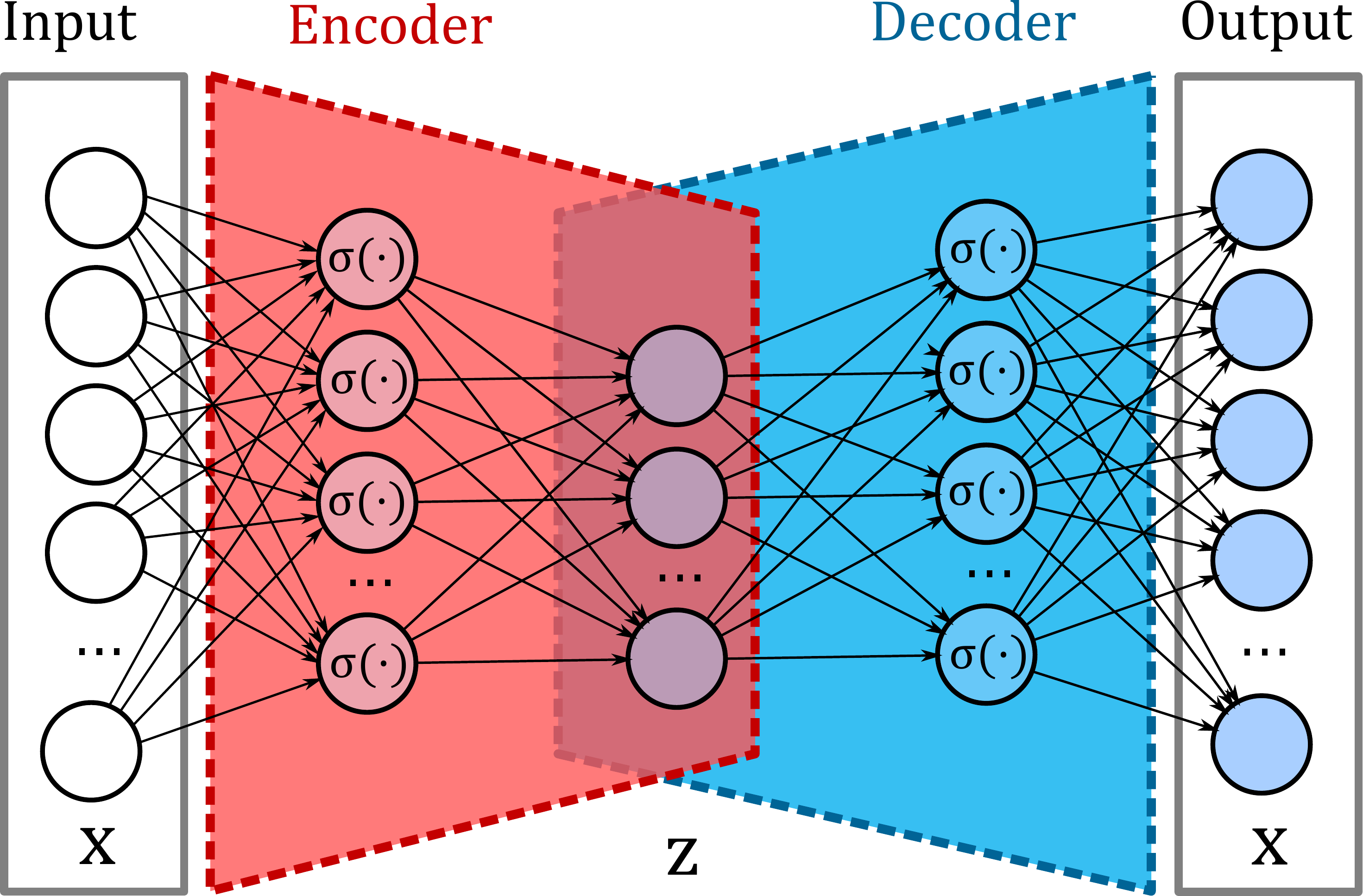}
\captionof{figure}{Autoencoder block with neural networks. A latent representation $\mathbf{z}$ is learned by the encoder and used as input of the decoder for recovering $\mathbf{x}$.}
\label{Autoencoder_picture}
\end{figure}

An autoencoder is a particular type of artificial neural network consisting of an encoding block which tries to learn a latent representation $\mathbf{z}$, typically in a lower-dimensional space, of the input variable $\mathbf{x}$, and a decoding block which reconstructs $\mathbf{x}$ at the output exploiting the code $\mathbf{z}$ (Fig. \ref{Autoencoder_picture}).

A typical learning formulation for deterministic autoencoders requires to solve
\begin{equation}
\label{AutoCost}
\theta_{\text{opt}} = \argmin_{\theta} \delta(\mathbf{x},G(F(\mathbf{x};\theta_1);\theta_2)),
\end{equation}
where $\delta$ is a distance measure, $\theta = (\theta_1,\theta_2)$, while $F$ and $G$ stand for the encoder and decoder function, respectively.
When $F$ and $G$ are linear functions, we get the Principal Component Analysis (PCA). Given a set/matrix $\mathbf{x}$ of $N$ $D$-dimensional samples $\mathbf{x}_i \in \mathbb{R}^D$, PCA sets the encoder as $F(\mathbf{x};\theta) = \mathbf{W}^T\mathbf{x}$ and the decoder as $G(\mathbf{z};\theta) = \mathbf{W}\mathbf{z}$ where $\mathbf{W}$ is the unknown parameter, a $D\times M$ matrix where $M$ is the dimension of the latent space. If $\delta$ is the quadratic loss function, then \eqref{AutoCost} can be rewritten as
\begin{equation}
\label{PCA}
\mathbf{W}_{\text{opt}} = \argmin_{\mathbf{W}} \sum_{i=1}^{N}{||\mathbf{x}_i-\mathbf{WW}^T \mathbf{x}_i||^2}.
\end{equation}
Let $\Sigma$ be the sample covariance matrix of $\mathbf{x}$, then $\mathbf{W}$ is given by the $M$ principal eigenvectors of $\Sigma$.
The geometric idea in PCA consists of finding a rotation of the domain space that aligns the principal axes of variance with the basis of the new representation latent space associated with $\mathbf{z}$.
Under a probabilistic model, the resulting transform is a vector $\mathbf{z}$ whose elements are mutually uncorrelated. PCA is broadly used also as a feature extractor, something which is very useful for the analysis of both the channel response and the noise in PLC, for instance.

Correlation is an indicator of linear dependence, but in most cases we are interested in representations where features have a different form of dependence. For example, in generative networks, Dinh \cite{Dinh2014},\cite{Dinh2016} proposed a non-linear deterministic transform of the data which maps them into a latent space of independent variables where the probability density results tractable.
Another interesting and suitable type of autoencoder for communications is the Denoising AutoEncoder (DAE) \cite{Vincent2008}. The idea is to train a machine in order to minimize the following denoising criterion:
\begin{equation}
\label{DAE}
\mathcal{L}_{\text{DAE}} = \mathbb{E}_{\mathbf{x}\sim \mathcal{D}}[\delta(\mathbf{x},G(F(\mathbf{\tilde{x}})))],
\end{equation}
where $\mathbf{\tilde{x}}$ is a stochastic corruption of $\mathbf{x}$. When we train a DAE using the expected quadratic loss and a corruption noise $\mathbf{\tilde{x}} = \mathbf{x}+\epsilon$ with $\epsilon \sim \mathcal{N}(0,\sigma I)$, Alain and Bengio \cite{Alain2014} proved that the autoencoder recovers properties of the training density $p(\mathbf{x})$.
They also showed that the DAE with a small noise corruption of variance $\sigma^2$ is similar to a Contractive AutoEncoder (CAE) with penalty coefficient $\lambda=\sigma^2$. The contractive autoencoder \cite{Rifai2011} is a particular form of regularized autoencoder which is trained in order to minimize the following reconstruction criterion:
\begin{equation}
\label{CAE}
\mathcal{L}_{\text{CAE}} = \mathbb{E}_{\mathbf{x}\sim \mathcal{D}}\biggl[\delta(\mathbf{x},G(F(\mathbf{x})))+\lambda \biggl| \frac{\partial F(\mathbf{x})}{\partial \mathbf{x}} \biggr|_F^2 \biggr],
\end{equation}
where $|\mathbf{A}|_F$ is the Frobenius norm. The idea behind CAE is that the regularization term attempts to make $F(\cdot)$ or $G(F(\cdot))$ as simple as possible, but at the same time the reconstruction error must be small. Typically, autoencoders find a low-dimensional representation of the input vector $\mathbf{x}$ at some intermediate level.

Now, in the domain of communications it has been proposed in \cite{OsheaIntro} to learn a robust representation $\mathbf{z}$ of the input $\mathbf{x}$ in order to overcome channel perturbations (noise, fading, distortion, etc.). In this way the transmitted signal can be recovered with small probability of error exploiting the \textit{redundancy} learned by the autoencoder. The findings in \cite{Alain2014} are somehow remarkable when applied in a communications framework because they assert that corrupting the transmitted signal with some form of noise can be beneficial for the autoencoder in the reconstruction's phase.

\subsubsection{Tools: Generative Networks}
\label{subsec:GAN}
Autoencoders can be studied also as generative models. Based on variational inference, Kingma and Welling introduced in \cite{Kingma2013} the concept of variational autoencoders. Denoted $\mathbf{z}$ as the latent variable of the observed value $\mathbf{x}$ for a parameter $\theta_1$, then $p(\mathbf{z|x};\theta_1)$ represents the intractable true posterior which can be approximated by a tractable one, $q(\mathbf{z|x};\phi)$, for a parameter $\phi$. A probabilistic encoder produces $q(\mathbf{z|x}; \phi)$ while a probabilistic decoder produces $p(\mathbf{x|z};\theta_2)$.
Rather than outputting the code $\mathbf{z}$, the encoder outputs parameters describing a distribution for each dimension of the latent space. When the prior is assumed to be Gaussian, $\mathbf{z}$ will consist of mean and variance. Tuning the parameters in the latent space, and passing the new latent samples through the decoder is a way to generate new data.

\begin{figure}
\centering
\includegraphics[scale = 0.27]{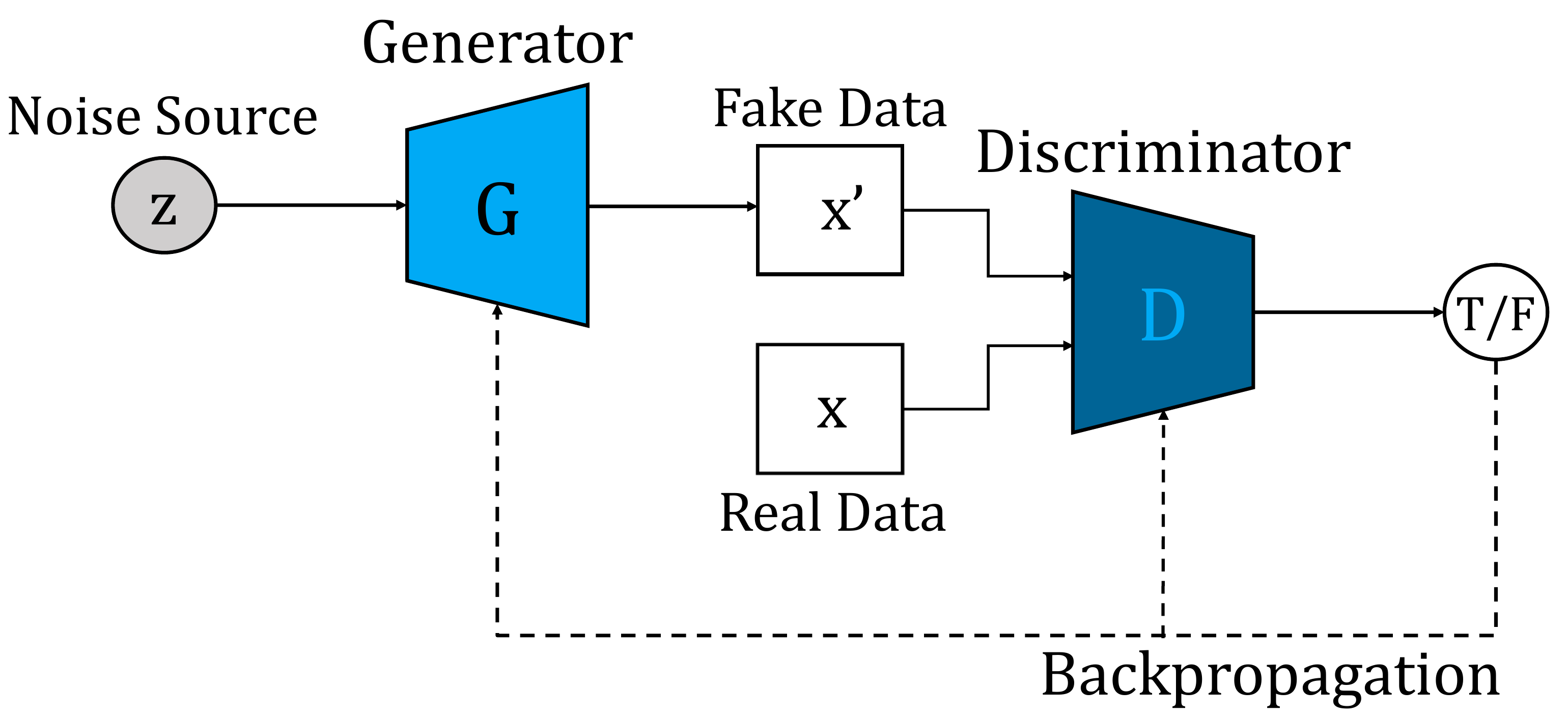}
\captionof{figure}{GAN framework in which the generator and discriminator are learned during the training process.}
\label{GAN_picture}
\end{figure}

Finally, in the generative models framework, it is worth mentioning Generative Adversarial Networks (GAN) proposed by Goodfellow et. al. in \cite{Goodfellow2014}. The main idea is to train a pair of networks in competition with each other (Fig. \ref{GAN_picture}): a generator model $G$ that captures the data distribution and a discriminator model $D$ that distinguishes if a sample is a true sample coming from real data rather than a fake one coming from data generated by $G$. The training procedure for $G$ is to maximize the probability of $D$ making a mistake. GANs can be thought as a minimax two-player game which will end when a Nash equilibrium point is reached. Given an input noise vector $\mathbf{y}$ with distribution  $p_{\text{noise}}(\mathbf{y})$, the map into data space is achieved through $G(\mathbf{y;\theta_{gen}})$. Defining the following value function
\begin{align}
V(G,D) = & \mathbb{E}_{\mathbf{x} \sim p_{\text{data}}(\mathbf{x})}[\log D(\mathbf{x})] \nonumber \\
& + \mathbb{E}_{\mathbf{y} \sim p_{\text{noise}}(\mathbf{y})}[1-\log D(G(\mathbf{y}))],
\label{GAN}
\end{align}
it was shown that the generator implicitly learns the true distribution since the equilibrium is reached when $p_{\text{gen}} = p_{\text{data}}$.

This generation technique has several applications in communication theory like stochastic channel modeling and channel coding \cite{OsheaGAN}. As we will discuss in the following (Sec. \ref{subsec:statistical_modeling}), it has also application in PLC due to the ability of generating synthetic data from a given distribution, thus, modeling for example the PLC channel response or noise.

\subsubsection{Application to Communications}
\label{subsubsec:appl}
Herein, we review some recent applications of unsupervised learning to communication systems for different layers of the protocol stack.
\begin{itemize}
\item \textit{Physical Layer}: Autoencoders for deep learning were firstly interpreted as a communication system in \cite{OsheaIntro}. The application of unsupervised learning to end-to-end communications can be divided according to the presence or not of an a priori knowledge of the channel model for the training part.
If the channel model is available, the usage of autoencoders with deep learning can help to design an optimal encoding/decoding procedure. For instance applications are found in optical fiber communications \cite{Karanov2018}, in AWGN channels with unreliable feedback \cite{Kim_Feedback2018}, sparse code multiple access schemes where deep neural networks are used to learn a decoding strategy to minimize bit-error rate \cite{Kim_SCMA2018}. When Channel State Information (CSI) is available, channel charting \cite{Studer2018} for localization exploiting CSI has been realized using autoencoders. If the channel model is unknown, several new approaches have been recently proposed. In order to design a radio communication system with autoencoders and to overcome the lack of channel knowledge, a two-phase training strategy was followed in \cite{Dorner2018}. For the same objective, \cite{Raj2018} and \cite{Aoudia2018} utilized stochastic perturbation and policy gradients techniques, respectively. Generative models can be used to generate samples of a given communication channel, in particular GANs have been exploited in \cite{OsheaGAN},\cite{YeGAN}.
\item \textit{Link and Medium Access Control Layer}: Spectrum sensing exploiting GANs and resource management for LTE were studied in \cite{Davaslioglu2018} and \cite{Challita2018}, respectively.
\item \textit{Network and Application Layers}: Clustering algorithms are the fundamental tools to be used in the network and application layers. Routing can be improved via clustering \cite{AbbasiClustering}, while in the application layer, identification of clustering communities, which is a social networks goal, has been considered in \cite{Abbe2014}.
\end{itemize}

\subsection{Other Learning Schemes}
\label{sec:other}
In the following sections, we briefly discuss Convolutional Neural Networks (CNNs), Recurrent Neural Networks (RNNs) and the Reinforcement Learning (RL) approach to offer an exhaustive overview of other ML tools that may have application in PLC.

\subsubsection{Convolutional Neural Networks}
Convolutional Neural Networks (CNNs) (Fig. \ref{OtherSchemes}.1) are able to capture the spatial and temporal dependencies of data, and for this reason they find application in image and document recognition \cite{LeCun1998}, medical image analysis \cite{Li2014}, natural language processing \cite{Collobert2008}, and more in general pattern recognition. CNNs are Multi Layer Perceptrons (MLP) (Sec. \ref{subsec:NN}) with a regularization approach since they consist of multiple convolutional layers to ensure the translation invariance characteristics. In particular, given an input data matrix $\mathbf{I}_i$, the feature map $\mathbf{F}_j$ is obtained as
\begin{equation}
\mathbf{F}_j = \sigma\biggl(\sum_{i=1}^{C}{\mathbf{I}_i \ast \mathbf{K}_{i,j}+\mathbf{B}_{j}}\biggr)
\end{equation}
namely, through the superposition of $C$ layers, e.g., $C=3$ for RGB images, each comprising a convolution between the input matrix $\mathbf{I}_{i}$ and a kernel matrix $\mathbf{K}_{i,j}$, plus an additive bias term $\mathbf{B}_j$, and a final application of non-linear activation function $\sigma(\cdot)$, typically, a \textit{sigmoid}, or \textit{tanh}, or \textit{ReLU}. Each set of kernel matrices represents a filter that extracts local features. To control the problem of over fitting, the dimension of data and features to be extracted is reduced by pooling layers. Finally, fully-connected layers are used to extract semantic information from features.

\subsubsection{Recurrent Neural Networks}
Recurrent Neural Networks (RNNs) are a class of supervised and unsupervised ML tools that are capable of representing time dependencies. Since they introduce the notion of time into the model, they have been successfully applied to tasks such as speech \cite{Graves2013} and handwriting recognition \cite{Graves2009}, health care \cite{Choi2016}, and machine translation \cite{DBahdanau2014}. Bearing in mind the architecture of the NNs presented in Sec. \ref{subsec:NN}, RNNs can be seen as feedforward NNs made of artificial neurons with one or more feedback loops (Fig. \ref{OtherSchemes}.2).

If we consider a simple RNN made of the input layer, an hidden layer, and the output layer, then the hidden layer or memory of the system $\mathbf{h}_t$, can be obtained as
\begin{equation}
\label{LayerRNN}
\mathbf{h}_t = \sigma(\mathbf{W}_{I,t}\cdot \mathbf{x}_{t}+\mathbf{W}_{H,t}\cdot \mathbf{h}_{t-1}+\mathbf{b}_t)
\end{equation}
where $\mathbf{x}_t$ is a real-valued input vector available at time $t$, $\sigma(\cdot)$ is the activation function, while $\mathbf{W}_{I,t}$, $\mathbf{W}_{H,t}$ and $\mathbf{b}_t$ are the weight matrix connecting the input to the hidden layer, and the weight matrix connecting the previous hidden layer to the current one and the biases, respectively. At time $t$, the hidden layer $\mathbf{h}_{t}$ is influenced by the current input $\mathbf{x}_{t}$ but also by the previous hidden state $\mathbf{h}_{t-1}$, such that the output $\mathbf{y}_{t}$ depends on the evolution over time of the hidden layer.

Several different architectures for RNNs have been proposed such as Long Short-Term Memory (LSTM) \cite{LSTM} and Bidirectional RNNs \cite{BRNN}. For further details and a wide survey on RNNs, the interested reader is referred to \cite{Lipton2015} and \cite{Salehinejad}, respectively.
\subsubsection{Reinforcement Learning}
Reinforcement Learning (RL) addresses the problem of an \textit{agent} learning to act in a dynamic \textit{environment} by finding the best sequence of actions that maximizes a \textit{reward} function. The basic idea is that the agent explores the interactive environment. According to the observation experience it gets, it changes his actions in order to receive higher rewards. RL finds several applications, from robots control \cite{Schulman2015}, to games such as Atari \cite{Mnih2013}, and Go \cite{AlphaGo}.

Basic RL can be modeled as a Markov Decision Process (MDP). Let $S_t$ be the observation (or state) provided to the agent at time $t$. The agent reacts by selecting an action $A_t$ to obtain from the environment the updated reward $R_{t+1}$, the discount $\gamma_{t+1}$, and the next state $S_{t+1}$.
In particular, the agent-environment interaction is formalized by a tuple $\langle \mathcal{S}, \mathcal{A}, T,r,\gamma\rangle$ where $\mathcal{S}$ is a finite set of states, $\mathcal{A}$ is a finite set of actions, $T(s,a,s')=P[S_{t+1}=s'|S_t = s, A_t = a]$ is the transition probability from state $s$ to state $s'$ under the action $a$, $r(s,a) = \mathbb{E}[R_{t+1}|S_t = s, A_t = a]$ is the reward function, and $\gamma \in [0,1]$ is a discount factor.
To find out which actions are good, the agent builds a \textit{policy}, i.e, a map $\pi : \mathcal{S} \times \mathcal{A} \to [0,1]$ that defines the probability of taking an action $a$ when the state is $s$. If we denote with $G_t = \sum_{k=0}^{\infty}{\biggl(\prod_{i=1}^{k}{\gamma_{t+i}}\biggr) R_{t+k+1}}$ the discount return, then the goal of the agent is to maximize the expected discount return (\textit{value}) $q^{\pi}(s,a) = \mathbb{E}_{\pi}[G_t|S_t = s, A_t = a]$, by finding a good policy $\pi(s,a)$. When the state or action sets are defined in high-dimensional spaces, the policy $\pi$ and the value $q$ can be represented by deep neural networks (Sec. \ref{subsec:NN}).

\begin{figure}
\centering
\includegraphics[scale = 0.30]{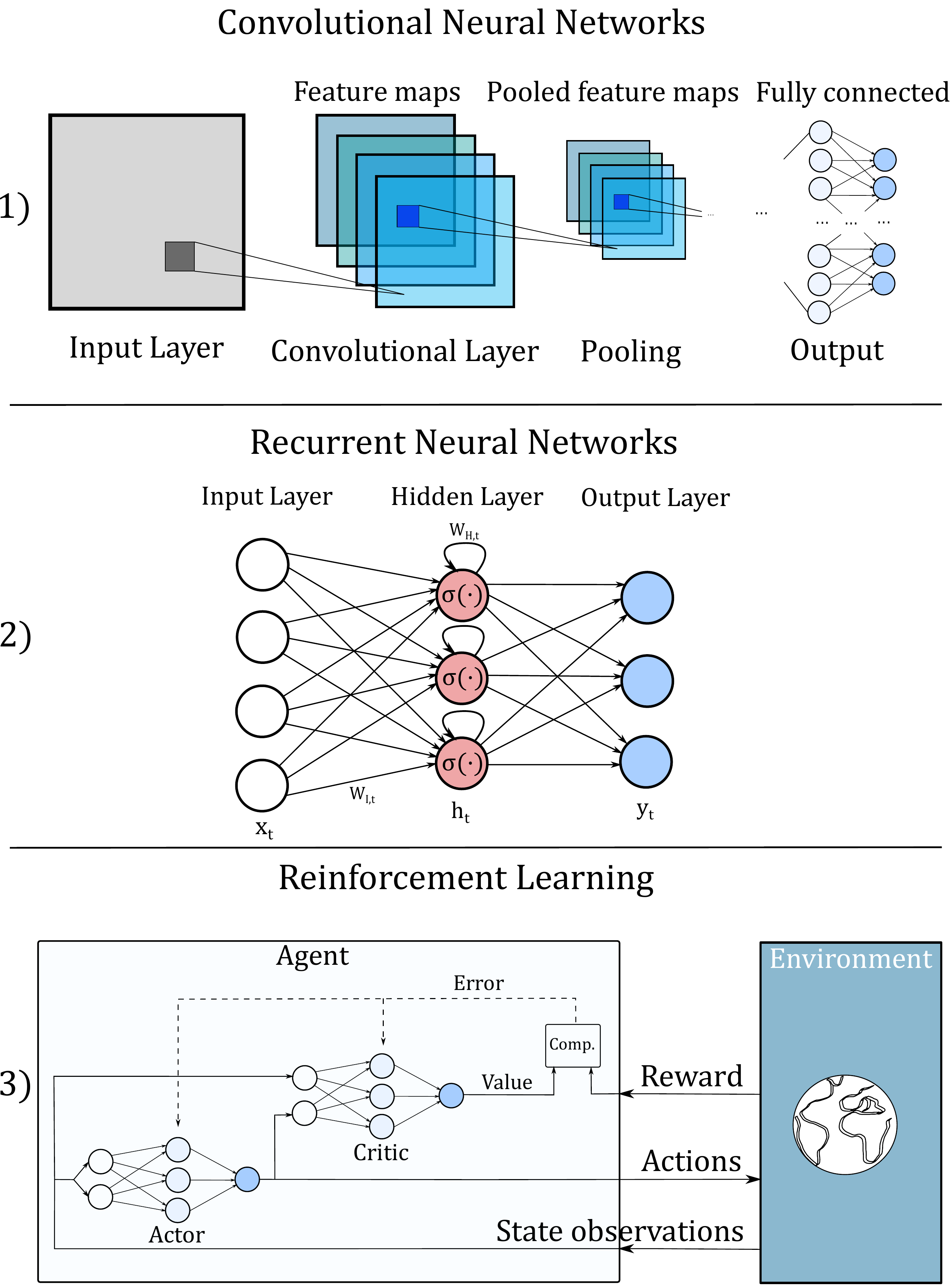}
\captionof{figure}{Graphical learning models of: 1) Convolutional Neural Network with one convolutional layer; 2) Recurrent Neural Network with simple feedback loop; 3) Reinforcement learning setting for actor-critic method.}
\label{OtherSchemes}
\end{figure}

With some more detail, there are essentially three different RL algorithms \cite{SuttonRL},\cite{ArulkumaranDRL}: a) policy based methods, when the agent, given the observation as input, optimizes the policy $\pi$ without using a value function $q$; b) value based methods, when the agent, given the observation and the action as inputs, learns a value function $q$; c) actor critic methods, where a \textit{critic} measures how good the action taken is (value-based), and an \textit{actor} controls the behaviour of the agent (policy-based) (Fig. \ref{OtherSchemes}.3).

\subsubsection{Application to Communications}
In \cite{Oshea2016}, CNNs naively extract features and use them for radio modulation recognition on time series of radio signals. A similar approach is adopted in \cite{Bitar} where CNNs automatically detect and identify frequency domain signatures for the IEEE 802.x standard compliant technologies.
RNNs have been firstly introduced as channel equalizers in \cite{Kechriotis}, while recently in \cite{RadfordRNN}, LSTM RNNs have been applied for network traffic anomaly detection.
Dynamic spectrum assignment in PLC has been investigated in \cite{7897112} under a RL approach. Due to the dynamic environment, RL finds several applications in communications and networking, mostly covered in \cite{RLCommSurvey}.

\section{ML for PLC Medium Characterization and Modeling}
In the following sections. We will mostly highlight the relevance and applications of the supervised and unsupervised ML tools described in Sec. \ref{sec:SL} and Sec. \ref{sec:UL} in this context. Reinforcement learning will also be briefly discussed in the context of resource allocation in PLC.  

We start from medium characterization since this is a key topic in PLC. It includes the analysis of the channel response (in time and frequency domain), the line impedance, the noise and interference \cite{PLCbookch2}.

\subsection{Statistical Learning}
\label{subsec:statistical_learning}
Up to date studies, have focused on the statistical characterization of the PLC channel and noise. For what matters the channel, datasets have been collected in several scenarios, such as in-home \cite{tlich2008indoor,5479916,5764406,6503661}, outdoor-access \cite{OPERADataset,5764389} and in-vehicle (cars, ships, planes) \cite{4913407,6525820,5764372,6201291,7445838}.
A great deal of knowledge has been acquired (mostly for the broad-band channel in the in-home scenario) and some consensus has been reached on the main characteristics. For instance, the frequency response and the average channel gain have a log-normal distribution \cite{galli2009a,tonello2014inhome}. Periodic channel time variations are exhibited especially in the narrow band spectrum, or for frequencies up to 10 MHz \cite{1650333}. Correlation is exhibited among the multi-conductor channels in MIMO setups \cite{6857317}.

Noise has a very complex nature in PLC since it comprises not only thermal noise components but also (and mostly) components generated by active loads and electronic circuits via direct coupling and conduction, e.g., harmonics and emissions from power converters, or inductive/EM coupling especially at high frequencies \cite{8360239}. Noise is extremely heterogeneous and temporal dependent. As such, its modeling is challenging. 
The existing description relies on a physical phenomena classification into stationary, cyclostationary, impulsive, and interference from radio transmissions that couple in the lines \cite{990732}. Only initial results for the description of the multi-conductor noise have been reported \cite{7052380,6812328,7150429,8245785}.

As in all communication systems, the accurate knowledge of the medium is a prerequisite for the design of a reliable and energy efficient communication technology. ML can bring new insights on the medium characterization. Firstly, it can be used to analyze and extract features. Then, to classify data from measurements.
The definition of the features that contain the information necessary to separate the data into classes is not a trivial step in the noise classification process.
Aiming at obtaining a statistical understanding of the medium, the features include energy, moments of multiple order, approximated entropy and correlation. A comprehensive list of features that can be considered, is reported in Tab. \ref{tab:features_description}. The knowledge about the PLC physical properties aids a lot to identify the most relevant features. For instance, the entropy is useful to recognize narrow band noise signals that are characterized by a sinusoidal behavior. The a priori knowledge that electromagnetic coupling effects among conductors may exist, suggests to study the correlation between noise traces in multiple conductor networks. 

\renewcommand{\arraystretch}{1.7} 
\begin{table*}
	\scriptsize 
	\centering
	\begin{minipage}{\textwidth}\centering
		\caption{Features description. }
		\begin{tabular}{ p{0.2cm}|p{1.1cm}|p{7cm}|p{6cm}}
			\toprule
			\textbf{ID} & \textbf{Feature name}  		& \textbf{Equation} & \textbf{Description}\\
			\midrule
			1&maxAbs &	$\max_j \{ \mid s_{j} \mid \}$	& maximum of the absolute values \\ \hline
			2&sum  & $\sum_{j=1}^N s_{j} $ & sum of the samples \\ \hline
			3&sum2 & $\sum_{j=1}^N s_{j}^2 $ & sum of the squared samples  \\ \hline
			4&std & $\sqrt{\frac{1}{N-1}\sum_{j=1}^{N}\mid s_j-\mu_i\mid ^2} $ & standard deviation of the samples\\ \hline
			5&skew &  $\frac{\frac{1}{N}\sum_{j=1}^{N}(s_j-\mu_i)^3}{\left(\sqrt{\frac{1}{N}\sum_{j=1}^{N}(s_i-\mu_i)^2}\right)^2} $  & skewness of the samples  \\ \hline
			6&kurt &  $\frac{\frac{1}{N}\sum_{j=1}^{N}(s_j-\mu_i)^4}{\left(\sqrt{\frac{1}{N}\sum_{j=1}^{N}(s_i-\mu_i)^2}\right)^2} $ & kurtosis of the samples  \\ \hline
			7&pears & $r_{CH1,2} = \frac{\sum_{j=1}^{N}(s_{CH1,i}-\mu_{CH1})(s_{CH2,i}-\mu_{CH2})}
			{\sqrt{\sum_{j=1}^{N}(s_{CH1,i}-\mu_{CH1})^2\sum_{j=1}^{N}(s_{CH2,i}-\mu_{CH2})^2}} $ & Pearson correlation of the samples  \\ \hline
			8&dist & $\|s_{CH1}-s_{CH2}\| = \sqrt{\sum_{j=1}^{N}\vert s_j\vert^2}$ & distance between data of channel 1 and 2  \\ \hline
			9&dCor & \cite{szekely2007} & distance correlation  \\ \hline
			10&ent & \cite{entropy_approx} & entropy approximation   \\ \hline
			11&diffEnt & \cite{entropy_approx} & entropy approximation of CH1-CH2  \\ \hline
			12&sumEnt & \cite{entropy_approx} & entropy approximation of CH1+CH2 \\ \hline
			13&fPeak & --- & find the number of samples over a certain voltage level  \\ \hline
			14&fEnFr & Exploiting the power spectral density (PSD) estimated via Burg's method. \cite{feature_extraction_book_healtcare} & find the energy of samples in a certain frequency range  \\ \hline
			15&fdist & --- & find the distance in samples between the two highest peaks  \\ \hline			
			16&corrStd & --- & standard deviation of the correlation of two channels  \\ \hline
			17&corrSkew & --- & skewness of the correlation of two channels  \\ \hline
			18&corrKurt & --- & kurtosis of the correlation of two channels  \\
		\end{tabular}
		\caption*{\\In this table $i$ and $j$ are respectively the index of the slot and the sample, $\mu_i$ is the samples mean of the $i^{th}$ slot, $s$ stands for the sample and $N$ is the length of the slot (in number of samples).}
		\label{tab:features_description}
		\noindent\makebox[\linewidth]{\rule{0.85\paperwidth}{0.4pt}} 
	\end{minipage}
\end{table*}

Classification tools can then be applied to determine the most prominent features and the significant number of classes.
An example is PCA (Sec. \ref{subsec:autoencoders}) analysis which is useful to reduce the dimensions of large datasets of variables and determine the most important features in the data.

\begin{figure}[t!]
	\centering
	\includegraphics[scale=0.7]{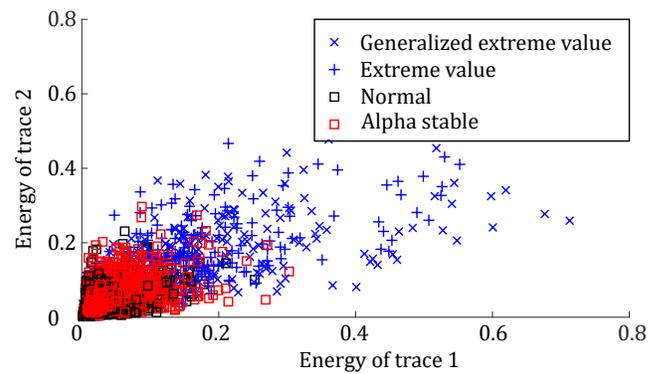}
	\caption{Example of noise data analyzed with SOM. The clusters are labeled with the associated PDF. The energy of the trace pairs 1 and 2 are normalized w.r.t. to their maximum.}%
	\label{fig:labled_pdf}
\end{figure}

Now, let us consider multi-conductor PLC noise. The exploitation of unsupervised learning allows the identification of noise classes. For this task, in \cite{righiniAutomaticClustering}, a library of features was created for a dataset of multi-conductor narrow band (NB) noise traces obtained via measurements in the band $3$-$500$ KHz in an office environment. Different methods were tested to extract the features (listed in Tab. \ref{tab:features_description}), and then provide a classification of the noise.  Features such as the distance correlation (\emph{dCor}), the Person correlation (\emph{pears}), or the differential entropy (\emph{diffEnt}) are useful to highlight the relations between the noise traces. SOM \cite{846731} provided the best results among the tested algorithms. Finally, for each class a statistical analysis of the associated traces was carried out.
An example of results obtained with this classification approach is shown in Fig. \ref{fig:labled_pdf}. Herein, the energy of the noise voltage trace on the first pair of conductors (neutral-ground, namely $V_{CH1}(t)$) is related to the energy of the noise trace on the second pair of conductors (neutral-phase, namely $V_{CH2}(t)$). The identified cluster PDFs are also reported in the legend of Fig. \ref{fig:labled_pdf}.

This data driven approach can classify and discover features in a more general way than the conventional approach that starts from a physical description of the phenomena generating noise, and essentially it partitions the noise into stationary, cylostationary synchronous/asynchronous with the mains frequency, and impulsive \cite{HanStoicaKaiser2016_1000056745}.

Another example is the analysis of the line impedance and the verification of whether it has a relation with the channel response \cite{6503661,TZheng19,dePianteChannelAdmittanceDependence}. This is a relevant question since should this be the case, then the forward link (transmitter-receiver) channel can be inferred from the measurement of the line impedance at the transmitter node \cite{dePianteChannelAdmittanceDependence}. In principle, the answer to this question could be obtained with a physical analysis of the lines and loads using Transmission Line (TL) and circuit theory. Some results following this approach have been recently obtained and proved that such a dependence does not exist if not at very low frequencies and for particular circumstances where the cables and wiring structures have conductors that do not have per-unit-length cross impedance terms \cite{dePianteChannelAdmittanceDependence}. However, one can object that the proof has been obtained under the assumption of transverse electromagnetic propagation for which TL holds true. Therefore, in general a non-linear, complex and time dependent relation may exist. Here is where ML comes into place: to discover dependencies.

Another example is the analysis of the dependence of the line impedance at the transmitter and at the receiver. Should such a dependence exist, then physical layer security techniques could be applied to exploit such common information for the generation of secure keys without the (insecure) exchange of information between sender and recipient \cite{XXXPasseriniPLCSecure}.

\subsection{Statistical-Synthetic Modeling}
\label{subsec:statistical_modeling}
Once the medium is characterized, next, modeling kicks in. A number of relevant results have been obtained for modeling the linear periodically time variant channel response \cite{5764406}. They include so called bottom-up approaches and top-down approaches. In the former case, TL theory is applied to a deterministic or statistical description of the network topology which includes the network graph, cable types, cable lengths, and loads \cite{tonello2011bottomup}. In the latter case, a mixed physical and phenomenological description of the channel response is given with the use of a parametric model derived from an abstract description of the link which accounts for multipath propagation, transmission-reflection effects, cable attenuation with length and frequency \cite{tonello2010bottomup,zimmermann2002a}.

\begin{figure}
	\includegraphics[scale=0.6]{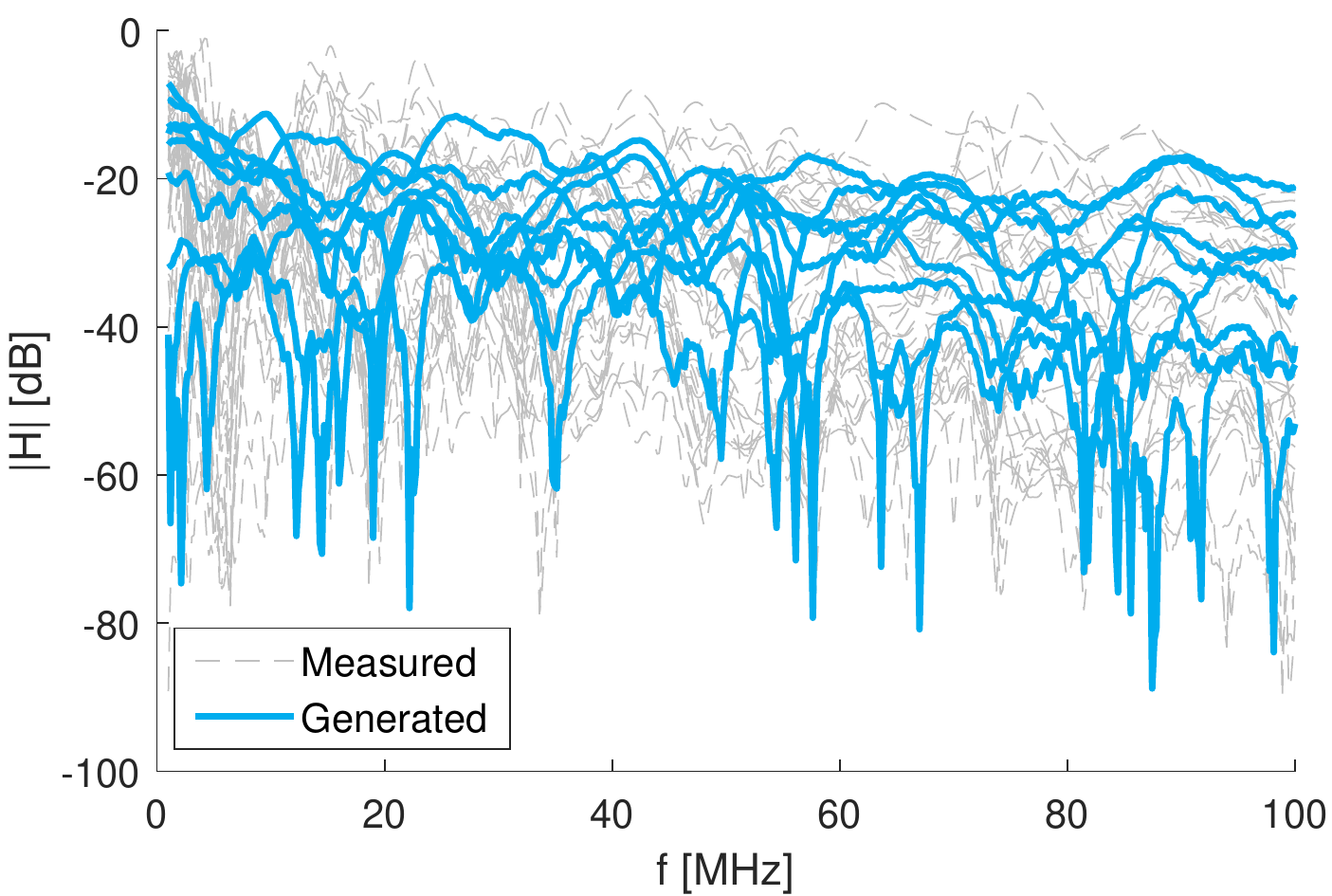}
	\caption{Broad-Band PLC channels measured and generated with GAN.}
	\label{fig:GAN-H}
\end{figure}

Deterministic or stochastic fitting of the model with data from measurements was done to derive either a deterministic model or a random channel model. More recently, it was observed that if we had an accurate statistical description of the channel response, then a purely phenomenological model could be derived \cite{7870625}. Essentially, such a model would be able to generate synthetic data that follow the observed statistics, totally abstracting from the physical interpretation of propagation of EM signals along the wires. This is essentially what ML approaches are nowadays trying to do when used as generators of \textit{fake} data in image and sound applications. Indeed the question is how to statistically analyze data and, in a more challenging way, how to generate synthetically such distributions. The work in \cite{7870625} proposed an approach which, with now an eye on ML techniques, can be further and significantly improved. For instance, GANs are an innovative tool to generate synthetic data following the same statistics of a given dataset. In particular, for communications, stochastic channel synthesis is a relevant topic since, for instance, new coding and modulation schemes require realistic channels to be effectively tested. Sec. \ref{subsec:GAN} describes the full methodology to design and train generative networks for this purpose. As an example, we report in Fig. \ref{fig:GAN-H} the result of using a GAN for the generation of the broad band PLC channel frequency response. The neural network architecture is trained with a dataset of 1000 measured single-input-single-output channel transfer functions with bandwidth of $100$ MHz obtained from measurements in the in-home environment \cite{tonello2014inhome}. Fig. \ref{fig:GAN-H} shows that the synthetically generated channel frequency responses look very consistent with the measured ones and span the large dynamic range of attenuation from $10$ to $90$ dB.

A similar approach can be adopted on other domains, for example, GANs along with RNNs (Sec. \ref{sec:other}) are promising tools to generate PLC synthetic noise while taking into account time dependence. In addition, another area where ML can provide help is modeling electro-magnetic-interference and mode conversions occurring along the medium on the signal \cite{PLCbookch3}. Radiation in PLC occurs at the network discontinuities and its radiation pattern depends on the length of each branch. The electromagnetic radiation of the cables can be accounted for in a per-unit-length equivalent circuit model using a series of resistances \cite{8017566}. An unsupervised optimization approach (using the techniques in Sec. \ref{sec:UL}) can be followed to obtain the optimal parameters. Alternatively, starting from an initial set of known parameters obtained for a given measurement dataset, supervised learning (using the techniques in Sec. \ref{sec:SL}) provides a methodology for the generalization and realization of a statistical channel generator that accounts for radiation effects. 

\section{ML for Physical Layer PLC}
PLC physical layer design \cite{PLCbookch5} encompasses the tasks of designing the channel coder, the modulator and the associated algorithms at the receiver side, which includes synchronization, channel estimation, detection/equalization, interference/noise mitigation, channel decoding. In addition, precoding at the transmitter side in the form of power allocation, bit allocation, precoding, and interference alignment are also \textit{tasks} classically relegated to the physical layer \cite{1716880,00927043,Rahman2015InterferenceAF}.

Physical layer PLC design was a vivid activity which brought innovation (partly derived from the results firstly obtained by the wireless communications community) in the form of:
\begin{itemize}
\item Analysis and optimization of concatenated codes, bit-interleaved codes, LDPC, Turbo codes, Fountain codes and associated decoding strategies taking into account the assumed peculiar amplitude statistics of noise, e.g., Middletone distributed \cite[Sec.~5.8]{PLCbookch5}.

\item Development of single carrier modulation (coded FSK, CDMA in first generation PLC devices), UWB modulation, multicarrier modulation (OFDM and filter bank modulation in second generation PLC devices) \cite[Sec.~5.2-5.5]{PLCbookch5}.

\item Precoding design taking into account the periodically time variant behavior of the medium so that the signal-to-noise ratio exhibits a periodic behavior over mains cycles \cite[Sec.~6]{PLCbookch6},\cite{1716880}.

\item MIMO techniques (also adopted by standards) and selection combining schemes including hybrid approaches that merge PLC links with radio links to improve diversity \cite{00927043,7436267}.
\end{itemize}

Although most of the above mentioned aspects are common to any communication system, the PLC channel specificities introduce several challenges, e.g., transmission is more band limited than in other systems, up-to-now a PSD constraint rather than a total power constraint has been imposed at the transmitter, spectrum management is relevant for coexistence, prediction of channel and noise is also relevant and may be enabled with new understanding of the medium. 
The question is whether ML can bring new \textit{lymph} in PLC PHY layer design. We envision a number of ways forward as detailed below.

Focusing at the receiver and at its tasks, ML can be used for estimation problems such as channel estimation and synchronization. For instance, assuming a certain parametric model of the channel, the parameters can be estimated via learning and inference. Similarly training of the equalizer can also be done via learning from field data.
More in general, the receiver task, i.e. data detection, can be seen as a classification problem. It can be solved with supervised learning, namely using NNs (Sec. \ref{subsec:NN}) or SVMs (Sec. \ref{subsec:SVM}), where the received signal is the input and the known transmitted bits are the output during the training process.

If we now look at the full communication system, a fascinating idea is to treat it as an autoencoder \cite{OsheaIntro} (Sec. \ref{subsec:autoencoders}). Let $s\in \mathcal{M}=\{1,2,\dots,M\}$ be a message out of $M$ possible ones and let $F:\mathcal{M}\mapsto \mathbb{R}^n$ be the transform realized by the encoding stage that maps the input message $s$ into the transmitted signal $\mathbf{x}=F(s)$.
This first block can be considered as the transmitter block which generates the waveform $\mathbf{x}=F(s)$. After it, the middle block of the autoencoder, learns the transform made by the channel and stochastically described by the conditional probability density function $p(\mathbf{y}|\mathbf{x})$ where $\mathbf{y}$ corresponds to the received signal. Lastly, the decoding block (receiver) takes $\mathbf{y}$ as input and estimates the transmitted message $s$ by computing the transform $G:\mathbb{R}^n \mapsto \mathcal{M}$. Each of these blocks can be implemented using NNs so that their chain constitutes a deep neural network which, trained end-to-end, reconstructs the input message at the output. Since the input is the message $s$ and the desired output is the same reconstructed message $s$, this leads to a classification task where the cross-entropy is a natural choice for the cost function. The end-to-end learning process enables the autoencoder to find a robust representation of the input signal which can lead to the identification of optimal coding and decoding schemes for stochastic channels. The autoencoder may find a solution that outperforms existing modulation and coding methods. In fact, the autoencoder approach works without any assumption on the channel and is able to recognize, in principle, its dynamics without the need to rely on the current view of the PLC channel as a linear periodically time-variant system \cite{5764406}.

\begin{figure}
	\centering
	\includegraphics[scale = 0.55]{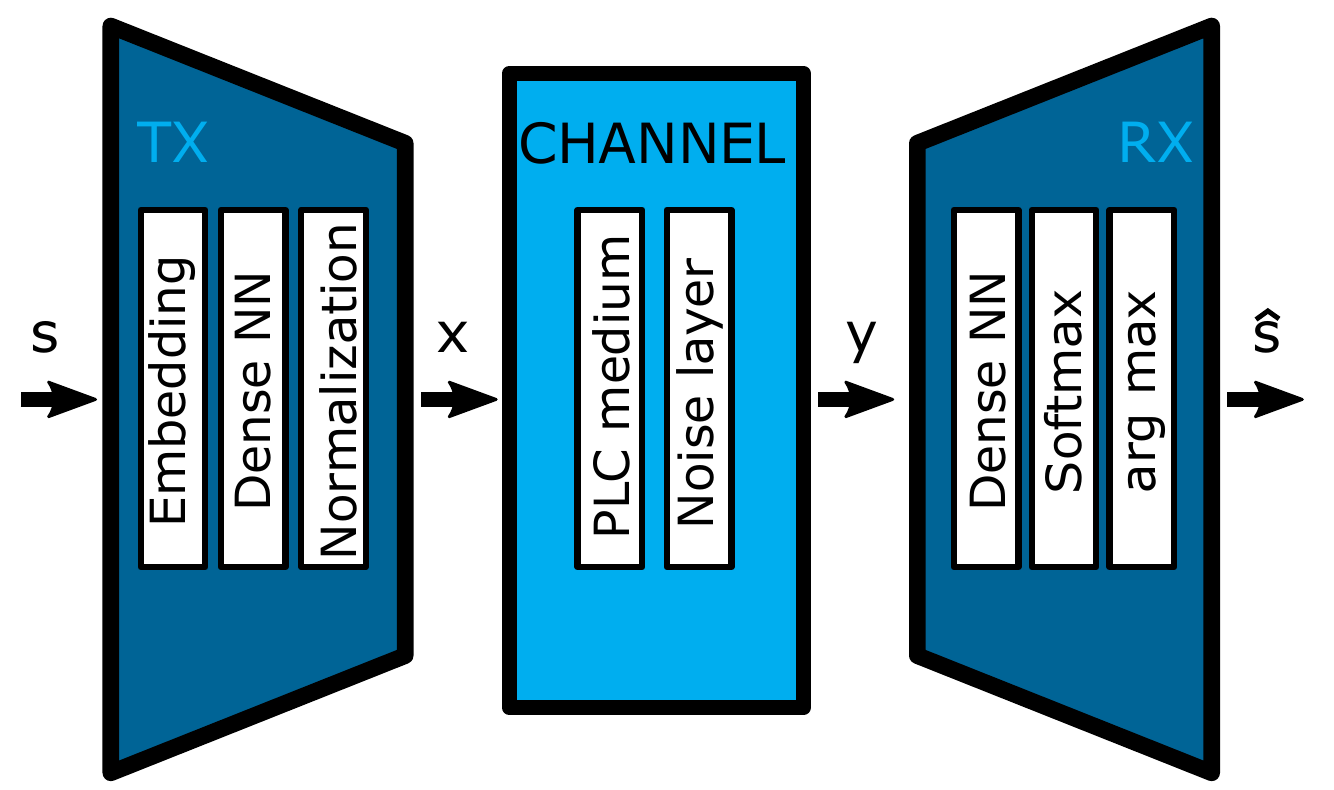}
	\captionof{figure}{Representation of a PLC system as an autoencoder.}
	\label{fig:autoencoder}
\end{figure}

\begin{figure}
	\centering
	\includegraphics[scale = 0.75]{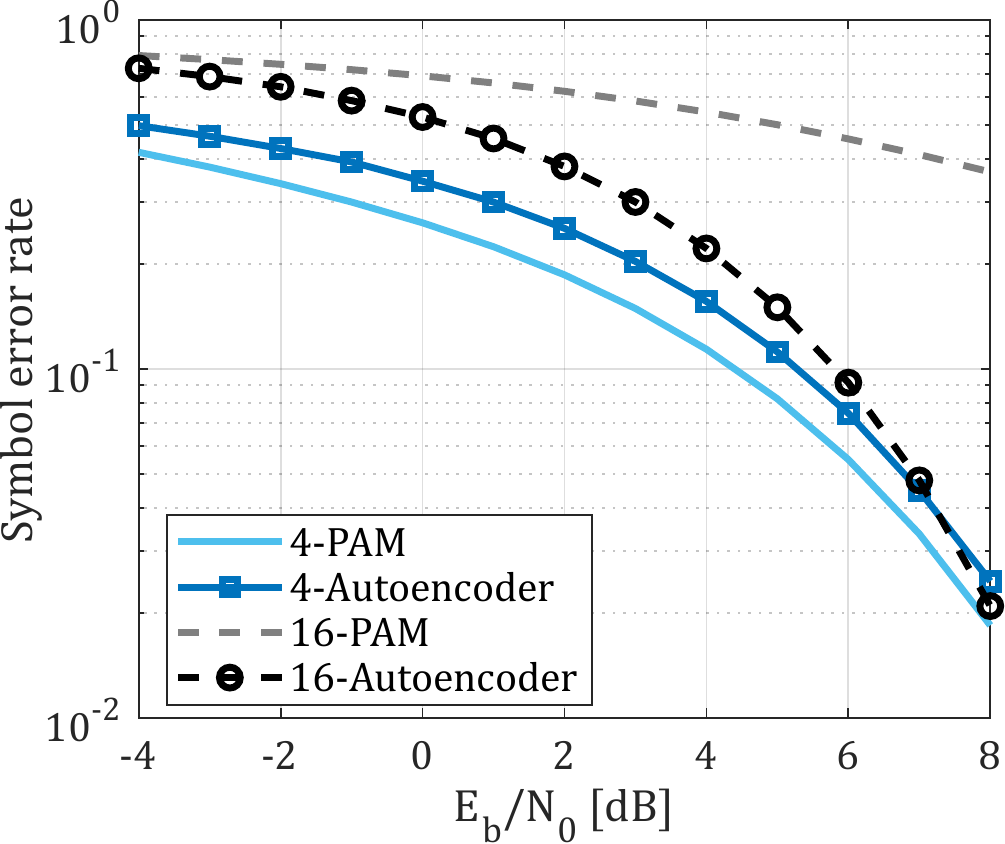}
	\captionof{figure}{SER versus $E_b/N_0$ for the autoencoder of Fig. \ref{fig:autoencoder} and the $M$-PAM modulation scheme.}
	\label{fig:SER}
\end{figure}

An autoencoder for application in PLC, schematically described in Fig. \ref{fig:autoencoder}, has been implemented using TensorFlow libraries \cite{TensorFlow}. The transmitter and receiver blocks were modeled as a cascade of multiple fully connected layers. The channel block was designed with a first layer that embeds the channel response (medium) and a second additive layer of Gaussian noise. An arbitrary channel response for the simulation was taken from the dataset described in \cite{tonello2014inhome} consisting of several channel realizations with bandwidth of $100$ MHz. The autoencoder takes as input symbols with cardinality $4$ and $16$ ($2$ or $4$ bits), respectively referred to as $4$-autoencoder, and $16$-autoencoder. The performance of the autoenconder (in terms of symbol error rate) has been compared to that attained by an impulsive $4$-PAM or $16$-PAM modulation scheme with an ideal matched filter receiver \cite{tonello2007wide}. The results are shown in Fig. \ref{fig:SER} as a function of the Signal-to-Noise Ratio (SNR) per bit.

It is interesting to notice that the 4-autoencoder performs 1 dB worse than the 4-PAM scheme in the considered SNR range which is also due to the fact that the 4-PAM scheme assumes perfect knowledge of the channel. However, by increasing the input signal cardinality the autoencoder manifests very good performance exceeding that of 16-PAM for all the considered SNRs. Furthermore, at a given SNR level we note an autoencoder cliff effect, i.e., a significant increase of the SER curve slope. This can be justified if we think that the autoencoder finds a more suitable non-linear coding and decoding scheme when it has a larger hyperspace to search in.

Another bunch of applications of ML for PLC is in the domain of optimization problems that arise at the transmitter side, namely, PHY layer resource allocation. Let us think for instance to the water filling problem that aims at distributing optimally power across the available spectrum \cite{PLCbookch6}. Instead of pursuing a physical model based approach, we may want to train a NN to provide the power allocation as the result of observing data from the field and defining a cost function which can be the bit-error rate or throughput. It should be noted that with a bottom-up approach this task can be solved by identifying the system state (channel response and noise), perform classification and estimation, and then dynamically determine the optimal allocation for the specific identified system state. However, a physical agnostic neural network can embed in one shot all these tasks and efficiently provide the solution. Of course, how to train the network and how complex the network has to be is a research question especially in PLC where the system state can have abrupt changes due to network topology changes, load changes, and noise generators changes. In addition, another approach based on reinforcement (Sec. \ref{sec:other})  learning can be exploited as done in \cite{7897112}.

\section{ML for MAC and Network Layer PLC}
Media access control and networking can also enjoy the use of ML. State-of-the-art PLC systems use contentious based media access techniques, i.e., CSMA, Aloha, flooding, as well as adaptive TDMA and FDMA \cite{PLCbookch6}. A universally optimal solution has not yet been found especially because of the heterogeneity of applications, services, traffic requirements, variability of medium conditions and network configurations. Let us think for instance of a typical automatic metering application in the access network managed by the concentrator at the MV-LV sub-station, in contrast to a PLC in-building network managed by cell coordinators, or a street light control application, in contrast to a PLC backhauling solution for 5G wireless micro-cells. Although chip sets and transceivers (both NB and BB) have been developed to serve as enabling communication devices for the plethora of above applications, and although they can provide point-to-point connectivity, the optimization of their performance in a massive networked environment is still a challenge. This is because multiple layers need to be optimized and so far limited data was available to learn how to exploit the wide flexibility of parameters at both the PHY and MAC layers. Things get even more complex when cooperation is required, i.e., when relaying and routing has to be implemented to offer full coverage in large massive networks \cite{5439083,D'Alessandro2012}. In brief, cross layer optimization, resource allocation, spectrum management and scheduling etc. are challenging tasks in PLC networks, as the results coming from field trials and large deployments are telling us. The dynamics of the channel and traffic, the abrupt network shut-downs and the need for fast reconfiguration, render these tasks extremely demanding if a layer by layer approach is considered, as traditionally done. This is because the traditional approach requires modeling each layer and function, which is prohibitively complex. Here is where ML comes into place.

ML and more in general artificial intelligence is envisioned as a relevant tool to develop a more abstract and synthetic model of the network behavior and to realize the network orchestration tasks \cite{7120043}. The network behavior has to be learned via the acquisition of real data sets where events, actions and results are labeled to create time series that can train optimization algorithms. In essence the orchestration of the network can be seen as a control problem where data observers serve as sensors, a global cost function is then formulated and minimized via learning from field data. To the best of our knowledge, efforts in applying ML techniques in routing problems have been carried out mostly for wireless networks and specifically Mobile Ad-hoc Networks (MANETs), where topology information is not available to nodes and the data-driven nature of the application poses significant challenges in terms of adaptation to run-time changes and scalability \cite{WSN_RL1,WSN_RL2,WSN_RL3,WSN_RL4}. Most studies also show that classical solutions, such as Bellman-Ford, Ad-hoc On-demand Distance Vector (AODV), Destination-Sequenced Distance-Vector (DSDV), Low-Energy Adaptive Clustering Hierarchy (LEACH), shortest path and the such, perform poorer than ML approaches when it comes to achievable throughput, end-to-end delay and time required to converge to a better solution when run-time changes occur. The most widespread approach to tackle these problems is reinforcement learning \cite{WSN_RL5}, where cost functions are defined in order to switch the behavior of a router based on environmental and link conditions (Sec. \ref{sec:other}). This is done by implementing a rewarding (or penalizing) system, which influences the probability of an element to make a specific choice, that keeps track of past states of the network and uses them to predict the possible future changes in the traffic load. The main disadvantage of RL is its computational demand, which weighs on every node of the network when implemented in a distributed kind of architecture. In the following, we present a different, computationally lighter approach, where supervised learning is used to synthetically train neural networks to implement a decentralized smart routing solution that is able to react to changes in PLC networks by using readily available information.

\begin{figure}
	\centering
	\includegraphics[scale = 0.25]{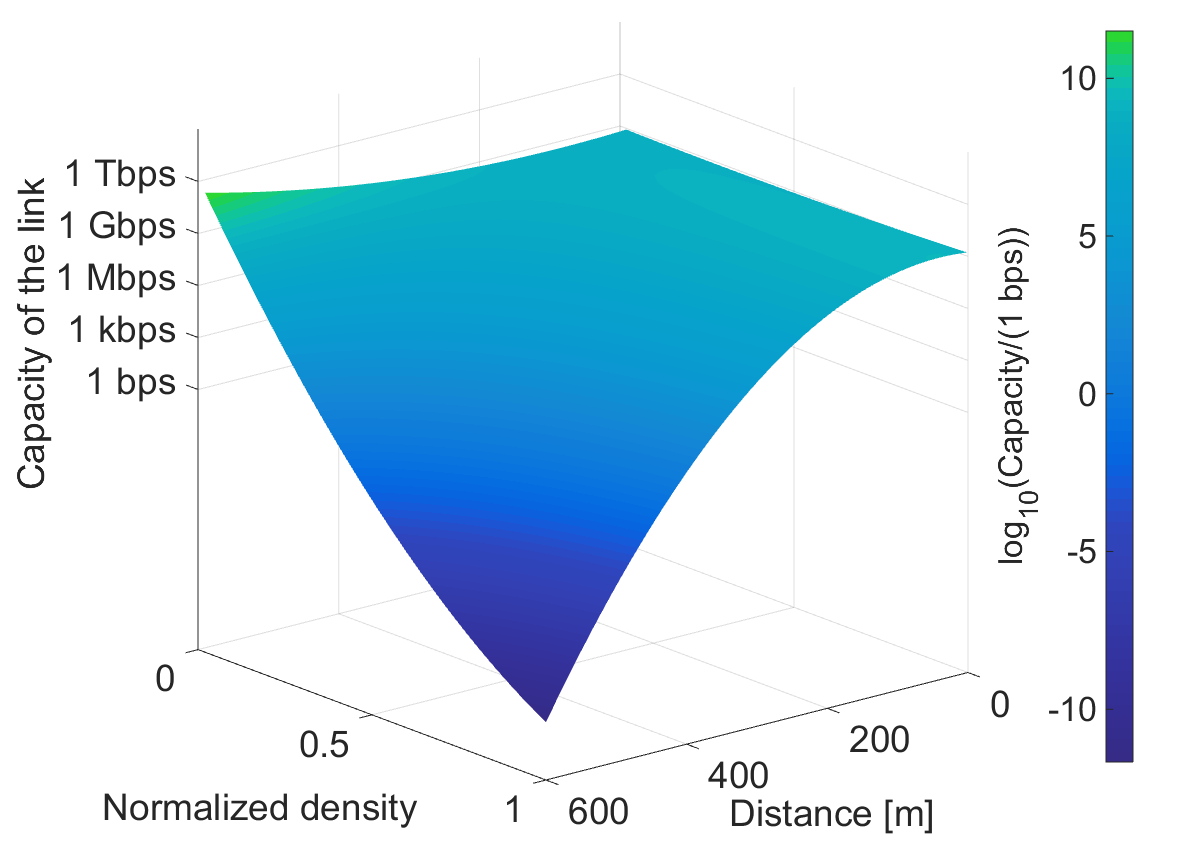}
	\captionof{figure}{Inferred average capacity of an end-to-end PLC link based on easily-observable topology parameters.}
	\label{fig:CapsVDD}
\end{figure}
In more detail, the envisioned application aims at using the power line distribution network as a fronthaul infrastructure for a small cell radio network. In \cite{ISPLC18Marcuzzi} the small cell radio network and the underlying power line infrastructure are brought together in a joint paradigm and the capacity of the PLC fronthaul is analyzed through a bottom-up emulation tool \cite{tonello2011bottomup,ICC2017Marcuzzi}. A regression approach is developed to tackle the problem of determining the capacity of an end-to-end power line link based solely on geometrical/topological properties of the network, for instance the density of nodes (radio cells) in the overall service area, and the distance between the communicating nodes. 

With the approach followed, it is possible to infer dependencies of the channel capacity and routing performance on a number of parameters (Fig. \ref{fig:CapsVDD}), e.g., topology type, density of nodes (branches), cable type, loads, electrical distance, etc. 
The channel model used to generate synthetic data is a TL model where the topology, loads, cable electrical characteristics etc. are included. It generates frequency selective channel responses for any pair of nodes in the network. Distance here refers to the length of the power line backbone that connects two considered nodes.
In detail, Fig. \ref{fig:CapsVDD} shows the capacity inferred through regression of two nodes pertaining to a PLC network when the communication is implemented in the broad-band spectrum (2-86 MHz). The curve shows an aberration for low densities and high distances: capacity does not actually increase with distance for low densities.
This is due to the fact that regression is implemented through a simulator that first deploys loads (small cells) on the territory and then tries to connect them in a way that power line is minimized. This means that for low densities, high distances are never achieved, thus the regression algorithm does not have data for that specific density-distance subset.

\begin{figure}
	\centering
	\includegraphics[scale = 0.21]{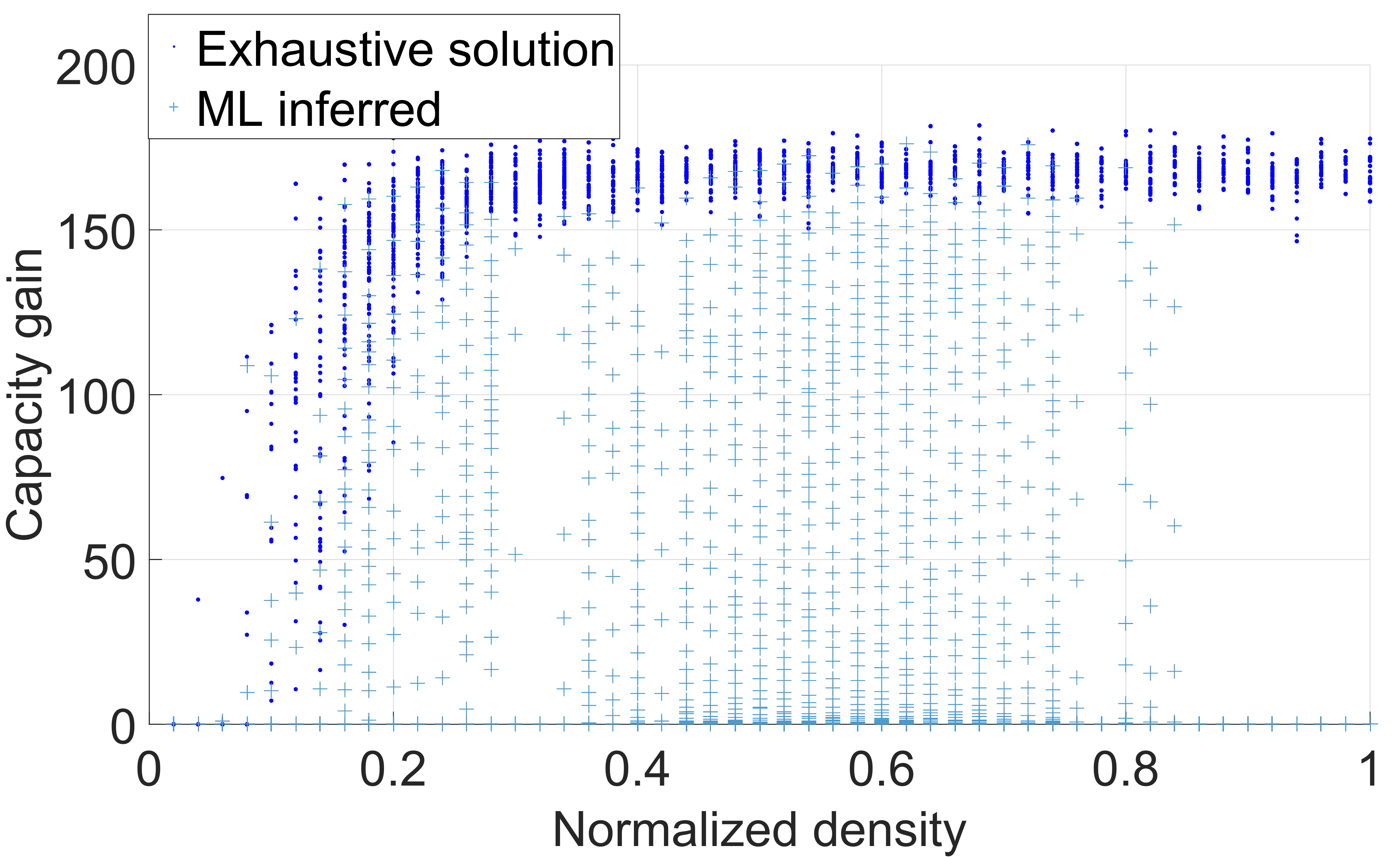}
	\captionof{figure}{Capacity gain using ML routing in a
PLC backhauling cellular network as a function of the radio cells density in the area.}
	\label{fig:capgain}
\end{figure}

The capacity inferred through regression enables a ML approach that can be used by nodes of the network to infer the best position of a global router to extend coverage and maximize capacity, given that the topology is known at every point. This information can be set by the manager of the power line infrastructure, otherwise, whenever a new node is added to the network, the topology can be reconstructed by a self-discovery mechanism. With every node being able to infer quickly and easily the capacity of the links in the network, a simple ML-based routing algorithm can easily increase capacity, as shown in Fig. \ref{fig:capgain}, where the capacity gained with ML routing is compared to the optimal solution, the latter achieved with full knowledge of topology and an exhaustive solution of the optimization problem.

\begin{figure}
	\centering
	\includegraphics[scale = 0.27]{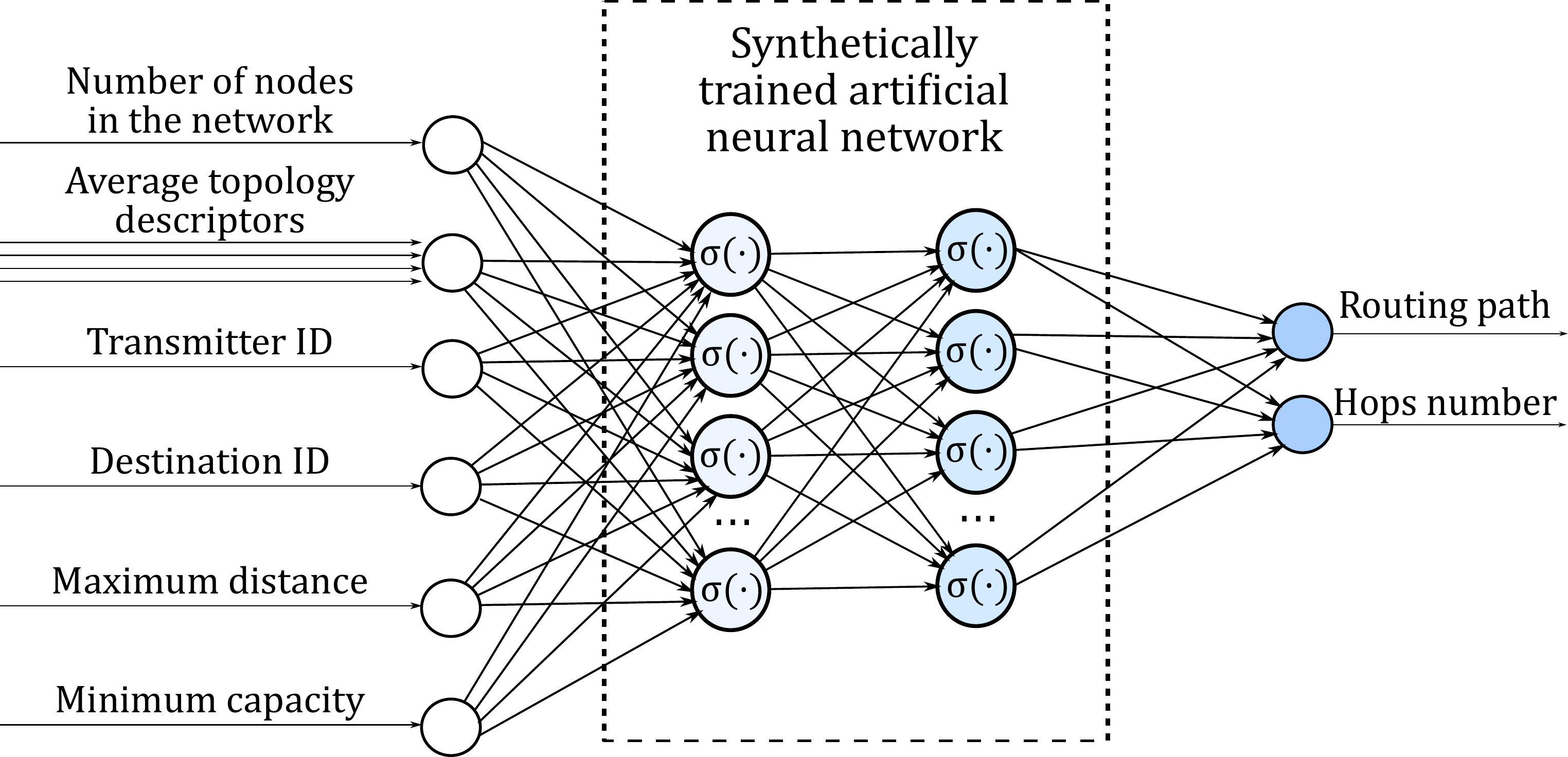}
	\captionof{figure}{NN used for determining routing.}
	\label{fig:NeuralNetworkPLC}
\end{figure}
\begin{figure}
	\centering
	\includegraphics[scale = 0.19]{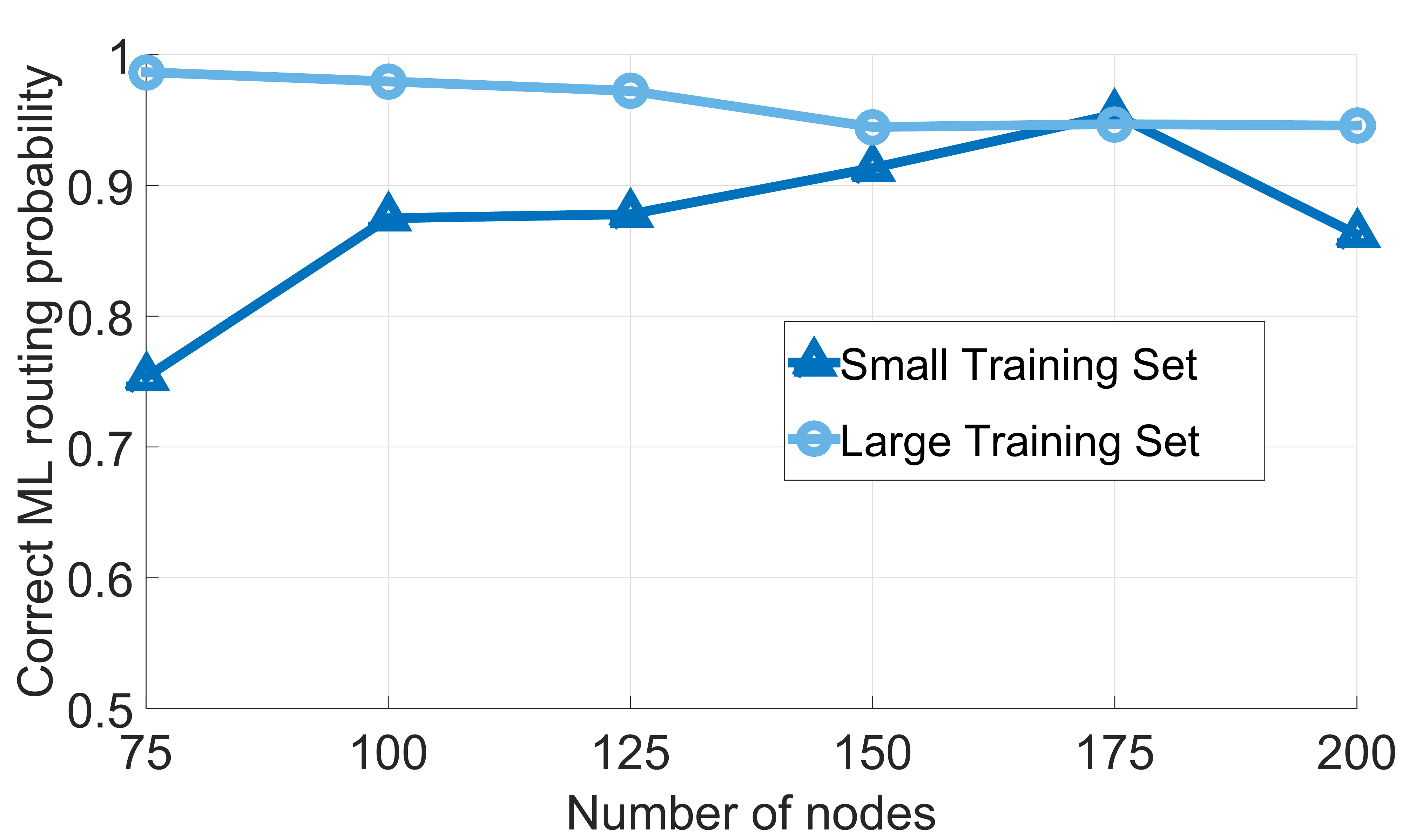}
	\captionof{figure}{Probability of determining the optimal routing scheme via NN as a function of the number of nodes and training set dimension in a metering PLC network.}
	\label{fig:AIPerf}
\end{figure}

On the other hand, we can also implement routing by considering the number of hops as the metric to be minimized, while imposing a constraint on the minimum capacity to be achieved. In addition, PLC can be used as an access network infrastructure for narrowband applications \cite{ISPLC2019Marcuzzi}. Sensor networks, which usually communicate via wireless technologies, can be supported by a PLC infrastructure where this solution has better penetration than a wireless one. A popular protocol for routing in these networks is the Lightweight On-demand Ad-hoc Distance-vector (LOADng) protocol developed by G3-PLC \cite{ISPLC2019Marcuzzi}. This protocol relies on information tables in order to connect all nodes in the network; topology is discovered by broadcasting route request packets, and routes are determined by minimizing a cost function. Asymmetric links, burst traffic and noise, as well as nodes with limited resources on power, computation and memory offer constraints that can be solved with lightweight neural networks. As shown in Fig. \ref{fig:NeuralNetworkPLC}, we can train our learning machine to devise the optimal number of routers and the best path for a message to reach any point in the network, both in downlink and uplink. The training set is herein generated synthetically by means of the aforementioned emulation tool, thus a big number of random network samples can be used to let the NN learn the underlying relations between topology and the optimal routing strategy. Specifically, the NN scheme accepts as inputs the source-destination IDs, the number of nodes in the network, the maximum distance between these nodes, the minimum capacity required to consider a link as functioning, and some other geometrical descriptors. The scheme outputs the number of routers required to achieve connectivity and the best path for connection. The basic idea of this implementation is that the nodes do not need to know the topology of the network, but only the few parameters referenced in Fig. \ref{fig:NeuralNetworkPLC}. Fig. \ref{fig:AIPerf} reports the percentage of correct guessed routing paths, i.e., the ratio of routing paths obtained by ML that are equal to the ones obtained with the exhaustive solution of the optimization problem and the perfect knowledge of the link capacity. The results show a high probability of success. Furthermore, it is worth noticing that the NN was trained with network topologies with a number of nodes ranging from 100 to 175. To prevent overfitting, an initial dropout layer \cite{Dropout} (drop rate $p=0.5$) was also added. The set used for testing the performance of the NN is nevertheless larger, i.e., networks with less than 100 or more than 175 nodes are simulated in order to assess the capabilities of our approach to find routing solutions for unobserved situations during the training stage. This explains the somewhat lower performance of ML routing w.r.t. the optimal routing for a number of nodes smaller than 100 and larger than 175.

In conclusion, based on the results presented, ML for networking in PLC is promising. Further investigations should be done: representative datasets have to be generated, supervised and unsupervised ML techniques have to be compared, and appropriate training algorithms have to be developed.

\section{Other Applications}
PLC is not a mere communication technology. The power grid is a fully interconnected system where all electrical phenomena affect the entire network. Moreover, it is influenced by external physical events. Hence, PLC nodes can be used to sense the grid itself as well as the surrounding environment. In other words, PLC transceivers can act as probes for grid diagnostics by analyzing the electromagnetic field and the data traffic in the PLC frequency bands \cite{AfricaTonello}. Possible applications of PLC for grid sensing are topology reconstruction \cite{lampe2013tomography,7797477} and anomalies detection, i.e., fault detection, cable aging, load identification \cite{8641473,8653266,7897106}. The former application enables grid operators to better the knowledge about the grid configuration, the status of switches and feeders which is not always complete especially in the LV part of the access network. The knowledge of the network topology is also important for networking aspects since it allows to locate nodes, and implement for instance geo-routing algorithms. The latter application enables grid operators to better monitor the grid status, malfunctions etc. which in turn enables predictive maintenance of the grid and therefore grants a better energy delivery service.
Some recent results have provided a better understanding of the propagation of high frequency signals (as PLC ones) in multi-conductor transmission lines \cite{8653266}. Essentially, it has been shown that we can relate the state of the grid, cables and loads to several electrical quantities that can be measured and that provide information to detect the presence of anomalies. The results offer an analytic framework described by a set of complex but extremely informative functions. Nonetheless, diagnosis of malfunctions ends up into an estimation problem or classification problem \cite{8641473}. In this context, \cite{7897106} has already provided results on the usage of ML for cable degradation detection, for instance.

To provide a numerical result, we consider the problem of detecting anomalies in a distribution grid by the analysis of high frequency PLC signals. In \cite{8653266} it was shown that for this objective, the line admittance at the nodes ($\mathbf{Y_{\text{in}}})$, the reflection coefficient ($\mathbf{\rho_{\text{in}}}$), and the transfer function ($\mathbf{H}$) between pairs of nodes can be exploited. Supervised learning (Sec. \ref{sec:SL}) can be very powerful here; once several sampled signals $\mathbf{x}$ are stored with the respective labels $y$, NNs or SVMs can learn the deterministic relationship $y = F(\mathbf{x})$ (which can be arbitrarily complex) and predict new unknown labels for different sensed signals. Here, we consider a simple neural network approach. A dataset was realized based on a single topology realization with $20$ nodes and average node length of $700$ m. Anomalies occurring on the grid have been added to obtain 10000 realizations of a perturbed grid, specifically, load impedances changes, concentrated faults and distributed faults according to the models described in \cite{8653266}.  
 

The training process was conducted using $50\%$ of the data, while testing was done using the other unbiased half part. We used only one hidden layer with $100$ neurons and a softmax layer to work with probabilities as output. Input signals are the ratios between the measurement of the electrical parameters $\mathbf{Y_{\text{in}}}, \mathbf{\rho_{\text{in}}}$ and $\mathbf{H}$ at a given time, and the reference value of the respective signal in the band $4.3-500$ kHz with $4.3$ kHz sampling frequency. The reference values were obtained when there was no anomaly. The output signals from the classifier are $4$ different class indicators according to the particular type of fault: 1. unperturbed, 2. load impedance change, 3. concentrated fault, 4. distributed fault. The obtained results are similar for all three types of electrical signals considered, so we present here only the accuracy values of the classifier when using the reflection coefficient. Furthermore, both the case of load impedances that stay constant (at $2$ k$\Omega$) or that are randomly variable are considered.

\begin{table}[]
\begin{tabular}{l|c|c|l|l}
\toprule
\multicolumn{2}{c|}{\multirow{2}{*}{\textbf{Accuracy}}} & \multicolumn{3}{c}{\textbf{Classes}}                                            \\ \cline{3-5}
\multicolumn{2}{c|}{}                                   & \multicolumn{1}{l|}{Fault Detection} & All 4 Classes & 3 Classes \\ 			
\midrule

\multirow{2}{*}{\textbf{\rotatebox[origin=c]{90}{Load} \textbf{\rotatebox[origin=c]{90}{Imped.}}}}    & Constant      & 87.8\%                               & 87.8\%                     & 100 \%       \\ \cline{2-5}
                                            & Variable   & 98.7\%                               & 89.2\%                    & 87.1\%        \\ \hline
\end{tabular}
\caption{Anomaly detection accuracy with constant and variable load impedances.}
\label{TableConf}
\end{table}

\begin{figure}
	\centering
	\includegraphics[scale = 0.33]{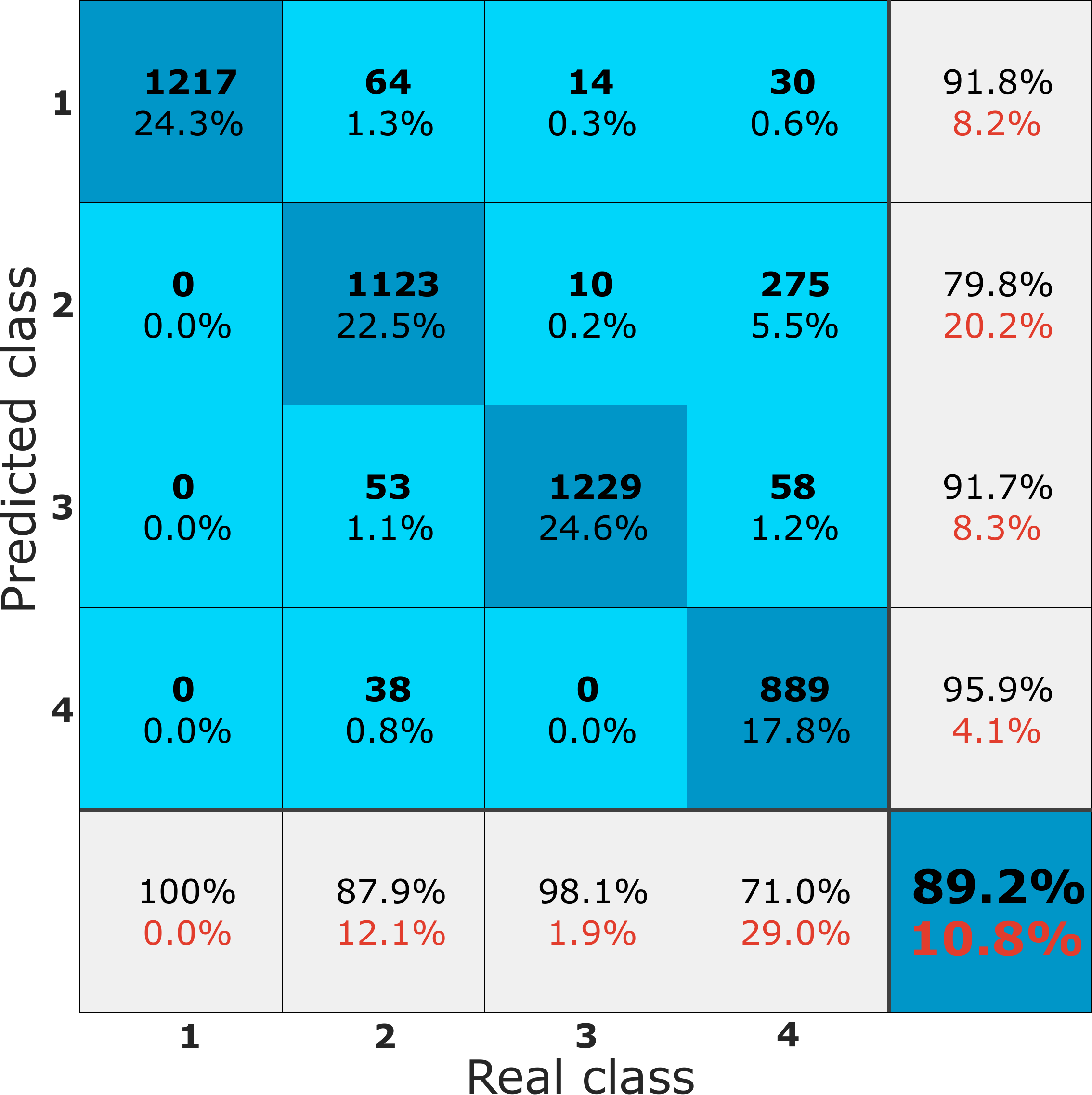}
	\captionof{figure}{Confusion matrix (accuracy of the prediction) for $4$-classes anomaly detection using the reflection coefficient as the main input signal.}
	\label{ConfusionExample}
\end{figure}

The first column of Tab. \ref{TableConf} reports the ability to detect the presence of a fault and this is achieved through an accuracy of $87.8\%$ when the load impedances are constant, and $98.7\%$ when they change, respectively. The second column shows the ability to detect and classify the type of anomaly: the average accuracy is $87.8\%$ when the load impedances are constant, and $89.2\%$ when they change, respectively. It is rather interesting to notice that, when the load impedances are constant, the neural network is not able to fully distinguish between the unperturbed network (class 1) and the distributed fault event (class 4). This suggests to remove from the data the last type of anomaly, and study the ability of the neural network to classify the remaining first $3$ classes. This is presented in the third column, where it is shown that the accuracy reaches $100\%$ when the load impedances are constant, and $87.1\%$ when they change, which means that most of the uncertainty is introduced by the last class. More details are offered by Fig. \ref{ConfusionExample}, which reports the confusion matrix of $4$ classes anomaly detection with variable loads. In this case, the confusion matrix shows a conflict between class 2 and class 4, in particular the NN erroneously tends to  predict (around $1/3$ of the times) class 2 when the correct class is instead the 4th. The results are promising and encourage further work for tuning the NN hyper-parameters.

\section{Conclusions}
This paper has provided an overview of ML as a signal/data processing tool to improve the knowledge and the performance of PLC systems. A clear distinction among deterministic and probabilistic ML formulations has been made.
Several new ideas of application of ML in PLC have been presented in the domain of medium characterization, statistical modeling, PHY layer, MAC layer and grid diagnostics. The paper has discussed what ML supervised and unsupervised tools can be used for the problem at hand. The concepts have been supported by numerical results using synthetic or real datasets. They have provided encouraging evidence that ML has a role in PLC. Despite its infancy, in this application domain, ML stimulates an all new era of research activities in PLC. 
Several future research directions have been envisioned, among which the exploitation of domain knowledge of PLC to better extract features via ML tools. For instance, the a priori physical knowledge of the channel (through parametric models) offers a guidance to better learn and characterize it. Furthermore, how to train, validate and test the machine (which is a core topic in ML), and how to compare the ML tools opens the door to several research endeavors. The starting point in PLC will be the development of representative data sets. This has to be followed by a statistical learning approach where the validity of results is measured in a probability framework.

\bibliographystyle{IEEEtran}
\bibliography{IEEEabrv,biblio}

\end{document}